\documentclass[12pt,fleqn]{article}

\usepackage{pgffor}
\usepackage{pdfpages}

\usepackage{amsmath,amsfonts,bm,amssymb}
\usepackage{amsthm}

 \usepackage{geometry}
\usepackage{rotating}
\usepackage{multirow}
\usepackage[greek,english]{babel}
\usepackage{latexsym}
\usepackage{wrapfig}
\usepackage{graphicx}
\usepackage{caption,subfig}
\graphicspath{{figures/}}
\usepackage{eurosym}
\usepackage{epstopdf}
\usepackage[authoryear]{natbib}
\usepackage{setspace}
\newcounter{pdfpages}

\begin{document}

\newgeometry{top=1cm, bottom=2cm}

\title{Do investors trade too much? \\A laboratory experiment}

\author{
Jo{\~a}o da Gama Batista$^\textrm{a}$,
Domenico Massaro$^\textrm{b,c}$\\
Jean-Philippe Bouchaud$^\textrm{d}$,
Damien Challet$^\textrm{a,e}$,
Cars Hommes$^\textrm{c,f}$
 }

%\date{\today}

\maketitle
\thispagestyle{empty}

%\vspace{0.2cm}

\small
\begin{center}
    $^\textrm{a}$ \emph{Laboratoire de Math\'ematiques et Informatique pour la Complexit\'e et les Syst\`emes, CentraleSup\'elec, Universit\'e Paris-Saclay}\\
    $^\textrm{b}$ \emph{Department of Economics and Finance, Universit\`{a} Cattolica del Sacro Cuore, Milano}\\%[4pt]
    $^\textrm{c}$ \emph{CeNDEF, Amsterdam School of Economics, University of Amsterdam}\\
    $^\textrm{d}$ \emph{Capital Fund Management, Paris}\\
 $^\textrm{e}$ \emph{Encelade Capital SA, Lausanne}\\
     $^\textrm{f}$ \emph{Tinbergen Institute}\\%[4pt]
\end{center}

\vspace{-0.15cm}

\maketitle

\vspace{-0.2cm} \small
%\begin{center}
 % First Draft: 11 September 2008\\
 % This Draft: 27 May 2009
%\end{center}

%\date{}

\begin{abstract}
% short (<= 100 words) abstract
%\singlespacing\footnotesize{We run an experiment to investigate the emergence of excess and synchronised trading activity leading to market crashes. Although the environment favours a buy-and-hold strategy, we observe that subjects trade too much, which is detrimental to their wealth given the implemented market impact. We find that preference for risk leads to higher activity rates and that price expectations are fully consistent with subjects' actions. In particular, trading subjects try to make profits by playing a \textit{buy low, sell high} strategy. Finally, we do not detect crashes driven by collective panic, but rather a weaker but significant synchronisation of market activity.}
% long abstract
\singlespacing\footnotesize{We run experimental asset markets to investigate the emergence of excess trading and the occurrence of synchronised trading activity leading to crashes in the artificial markets. The market environment favours early investment in the risky asset and no posterior trading, i.e. a buy-and-hold strategy with a most probable return of over $600\%$. We observe that subjects trade too much, and due to the market impact that we explicitly implement, this is detrimental to their wealth. The asset market experiment was followed by risk aversion measurement. We find that preference for risk systematically leads to higher activity rates (and lower final wealth). We also measure subjects' expectations of future prices and find that their actions are fully consistent with their expectations. In particular, trading subjects try to beat the market and make profits by playing a \textit{buy low, sell high} strategy. Finally, we have not detected any major market crash driven by collective panic modes, but rather a weaker but significant tendency of traders to synchronise their entry and exit points in the market.}
\end{abstract}
\medskip


%\medskip


%\newpage

%\footnotesize{\textbf{JEL codes:} }
%\medskip

%\singlespacing\footnotesize{\textbf{Keywords:} }
%\medskip
\bigskip
\singlespacing\footnotesize{\textbf{Acknowledgments:} This work was partially financed by the EU ``CRISIS'' project (grant number: 
FP7-ICT-2011-7-288501-CRISIS), Funda\c{c}{\~a}o para a Ci\^{e}ncia e Tecnologia, and the Ministry of Education, Universities and Research of Italy (MIUR), program \textit{SIR} (grant n. RBSI144KWH).}

\newpage

\normalsize \onehalfspacing

\restoregeometry

\section{Introduction}\label{intro}

Financial bubbles and crises are potent reminders of how far investors' 
behaviour may deviate from perfect rationality. Many behavioural biases of individual investors are now
well documented, such as the propensity for trend following or extrapolative 
expectations \citep{greenwood2014expectations}, herding behavior \citep{cipriani2014estimating}, disposition effect \citep{grinblatt2001investors}, home bias \citep{solnik2012global}, over and underreaction to news \citep{barber2008all}; see \cite{barberis2003survey} and \cite{Barber2013behavior} for comprehensive overviews. 

A well established fact about individual trading behaviour which is in stark contrast with the predictions of rational models is the tendency of individual investors to trade too much \citep{odean1999tradetoomuch}. Many investors trade actively, speculatively, and
to their detriment. \cite{odean1999tradetoomuch}, \cite{barber2000trading}, \cite{odeanboys} and \cite{barber2009just} among others show that the average return of individual investors is well below
the return of standard benchmarks and that the more active traders usually perform worse on average. In other words, these investors would do a lot better if they traded less. Moreover, as noted in \cite{Barber2013behavior}, individual investors make systematic, not random,
buying and selling decisions.

Based on the empirical evidence mentioned above, the goal of this paper is to study the emergence of excess trading in an experimental financial market where trading is clearly detrimental for investors' wealth. We aim to gain a deeper understanding not only of why agents trade, but also how correlated is their activity. In particular, we want to study whether synchronisation of trading activity leads to unstable market behaviour, such as crashes driven by panic, herding and cascade effects. 

Market laboratory experiments now have a rather long history. The most influential paradigm for multi-period laboratory asset markets was
developed in \cite{smith1988bubbles}. The asset traded in their experiment has a known finite
life span and pays a stochastic dividend at the end of each period. The fundamental
value of the asset falls deterministically over time, and lacking a terminal value the asset expires worthless. A salient result is that asset prices in the experimental markets follow a
``bubble and crash'' pattern which is similar to speculative bubbles observed in real world
markets.\footnote{In fact, \cite{Smith2010Theory} blames a failure of backward induction
for the existence of bubbles in these simple experimental markets and he suggests that \textit{``price bubbles
were a consequence of \dots homegrown expectations of prices rising''}. \cite{oechssler2010searching} shares this view and argues that \textit{``backward induction is only useful when there
is a finite number of periods which most asset markets don't have. Subjects
are told that they trade assets on a market so they probably expect to see
something similar to what they see on real markets: stochastic processes
with increasing or at least constant trend in most cases.''} See e.g.~\cite{noussair2010peaks}, \cite{giusti2012eliminating}, \cite{Breaban2014fundamental} and \cite{stockl2015multi} for previous experimental studies on markets with (partly) increasing fundamental values.}  This seminal work has spawned a large number of replications and follow-ups, see \cite{Palan2013review} for an extensive overview. 

The present study belongs to the above tradition of experimental markets but implements a different market mechanism. Instead of trading in a continuous double auction, subjects can submit buying and selling orders, executed by a market maker, for a fictitious asset that increases in value at a known average rate and has an indefinite horizon.

One particularly interesting feature of our laboratory market is that we model and implement
price impact, i.e.~the fact that the very action of agents modifies the
price trajectory. This is now believed to be a crucial aspect of real financial 
markets, which may lead to feedback loops and market instabilities \citep{Bouchaud2009,cont2014fire,caccioli2014stability,lillo2003econophysics,lillo2004long}.
What is of particular interest in our experiment is that
excess trading significantly impacts the price trajectory and is strongly 
detrimental to the wealth of our economic agents. In other words,
unwarranted individual decisions can lead to a substantial loss of collective 
welfare, when mediated by the mechanics of financial markets.

One advantage of using a controlled laboratory environment is that it allows us to directly measure individual characteristics and relate them in a systematic way to trading activity. Previous empirical studies show that individual risk attitude could have an impact on trading in asset markets. For example, \cite{robin2012bubbles} and \cite{fellner2007risk} find that
risk-aversion leads to smaller bubbles and less trade in asset market. Moreover, \cite{Keller2006Investing} did a mail survey and found that financial risk tolerance is a
predictor for the willingness to engage in asset markets. In the light of the aforementioned empirical evidence, we measure subjects' risk aversion using a standard \cite{holt2002risk} procedure and link it to individuals' trading activity.

We also elicit individual price expectations in order to better understand what leads to investors' decision of engaging actively in trading activity as well as being inactive in the market. Since we collect both data on individual trading decisions as well as price expectations, we are able to give a consistent picture of activity and inactivity as a consequence of price return expectations. See also \cite{smith1988bubbles}, \cite{hommes2005coordination,hommes2008expectations} and \cite{haruvy2007traders} among others for studies on the role of expectations in generating bubbles and crashes in experimental asset markets. 

Our findings can be summarised as follows. Although the market environment clearly favours a buy-and-hold strategy (see below), we observe that our subjects engage in excessive trading activity, which is both individually and collectively 
detrimental, since the negative impact of sellers reduces the price of our artificial asset. 
When the experiment is immediately repeated with the same subjects, we see a significant 
improvement of the collective performance, which is however still
substantially lower than the (optimal) buy-and-hold strategy. 

Moreover, although our subjects are physically separated and cannot communicate, we have 
seen that a significant amount of synchronisation takes place in the decision 
process, that can therefore only be mediated by the price trajectory itself. This resonates with what happens in real financial markets, where price changes 
themselves appear to be interpreted as news, leading to self-reflexivity and 
potentially unstable feedback loops (see e.g. \citet{Bouchaud2009,hommes2013reflexivity}). In fact, our experimental setting was such 
that panic and crashes were possible but this did not happen. Although we observed a significant level of synchronisation, no cascades or ``fire sales'' 
effects could be detected for our particular choice of parameters. 

Consistently with the empirical evidence mentioned above, we find that risk loving attitudes lead to higher trading activity and this is detrimental to individual wealth.

Finally, subjects seem to have a desire to trade actively, motivated by a willingness to ``beat the market'', as revealed by the analysis of individual price expectations. Most subjects engage in trading activity trying to {\it buy low and sell high}, while the decisions of being inactive and hold cash (shares) depends on whether they expect price returns lower (higher) than average.

The outline of this paper is as follows. Section 
\ref{sec:exp_design} describes the experimental motivation and set-up. The 
precise instructions given to
our subjects are detailed in Appendix \ref{sec:instructions}. Section
\ref{sec:rational_bench} explains the theoretical benchmark to which we want to 
compare the experimental results. In
particular, we show that (risk-neutral) rational agents should favour, in the present market setting, a 
buy-and-hold strategy. We then summarize our main results in Section
\ref{sec:results}, which includes a refined statistical analysis of agents' trading activity in Section \ref{ssec:global}. In Section \ref{ssec:risk} we relate individual risk attitude to activity rate and final wealth, while in Section \ref{ssec:predictions} we link price forecasts and trading behaviour. Section \ref{sec:conclusion} offers our 
conclusions, open questions, and future experiments.

\section{Experimental Design}\label{sec:exp_design}
The basic idea of our experiment is to propose to subjects a simple investment 
``game'' where they can use the cash they are given at the beginning of the experiment to invest in
a fictitious asset that will -- they are told -- increase in value at an 
average rate of $m=2 \%$ per period. The asset ``lives'' for an indefinite horizon, i.e.~the game may stop randomly at each time step with probability $p=0.01$. The
game is thus expected to last around $100$ time steps.
Subjects are also informed that random shocks impact the price of the asset in each period. In the absence of trade, price dynamics are described by
\begin{equation}\label{eq:priceNOtrade}
p_{t+1} = p_t \cdot \exp(m+s\eta_t),
\end{equation}
where $m=2\%$, $\eta_t$ is a noise term drawn from a Student's t-distribution with 3 degrees of freedom and unit variance,\footnote{Mathematically, the 
average of the exponential of a Student distribution is infinite, because of rare, but extreme values of $\eta_t$. In order not to
have to deal with this spurious problem, we impose a cut-off beyond $|\eta|=10$, with no material influence on the following discussion.} as commonly observed in financial markets  \citep{gopikrishnan1998inverse, jondeau2003, bouchaud2003theory}, and $s$ is a constant that sets the actual 
amplitude of the noisy contribution to the evolution of the price (i.e. the 
price
volatility) and is chosen to be $s=10\%$. These numbers correspond roughly to the 
average return and the volatility of a stock index over a quarter. Therefore, in terms of returns and volatility,
one time step in our experiment roughly corresponds to three months in a real 
market, and 100 steps to 25 years.

If an amount $w_0$ is 
invested in the asset at time $t=0$, the wealth of the inactive investor will accrue
to
\begin{equation}\label{eq:price}
	w_T = w_0 \exp \left[ m T + \xi s \sqrt{T} \right]
\end{equation}
at time $T$, where the second term of the exponential implies random fluctuations of root mean square (RMS) $s$ per period with $\xi$ defined as a noise term with zero mean and unit variance. The 
numerical value of the term in the exponential is therefore
equal to $2 \pm \xi$ for $T=100$, leading to a substantial most probable profit of $e^2 - 
1 \approx 640 \%$. As we shall show below, the \textit{fully rational} decision in the presence of risk-neutral agents
is to buy and hold the asset until the game ends; for the students
participating in the experiment, the most probable gain would represent roughly 
$\text{EUR } 160$, a very significant reward for spending two hours in the
lab. In other words, the financial motivation to ``do the right thing'' is 
voluntarily strong.

In order to make the experiment more interesting, and trading even more 
unfavourable, the asset price trajectory is made to react to the subjects 
decisions, in a way that mimics {\it market impact} in real financial markets. 
The idea is that while a buying trade 
pushes the price up, a selling trade pushes the 
price down (for a short review on price impact, see e.g. \citet{Encyclopedia}). As an agent submits a (large) buying 
or selling order at time $t$, the price $p_{t+1}$ at which the transaction is 
going to be fully executed is (a) not known to him at time $t$ and (b) 
adversely impacted by the very order that is executed. It is made very clear to the subjects that their transaction orders will be executed at the impacted price, 
meaning that the impact will amount for them as a cost. This should therefore be a strong incentive not to trade. 

The price updating rule described in Eq.~\eqref{eq:priceNOtrade} is easily modified to include price impact and now reads\footnote{We interpret $I_t$ as the permanent component of the 
price impact, which we assume to be non-zero, meaning that even random trades do affect the long term trajectory of the price. This is clearly at odds with the efficient market theory, 
within which the price impact of uninformed trades should be zero. We tend to believe that the anchoring to the ``fundamental price'' in real markets is very weak and is only relevant 
on very long time scales \citep{Thaler85}, so that the assumption made here is of relevance for understanding financial markets. See the discussion in \citep{BFL09,Donier15}}
\begin{equation}\label{eq:price_update}
	p_{t+1}=p_{t}\cdot\mathrm{exp}\left(m+s\eta_{t}+I_{t}\right).
\end{equation}
\noindent The term $I_{t}$ is the price impact 
caused by all the orders submitted at time $t$, which we model as:
\begin{equation}\label{eq:impact}
	I_{t}=\frac{N_{t}}{N}\frac{B_{t}-S_{t}}{B_{t}+S_{t}},
\end{equation}
where $N_{t}$ is the number of subjects who submitted an order at time $t$ and $N$ 
is the total number of subjects in a given market session (i.e. the ``depth'' of our market). 
$B_{t}$ and $S_{t}$, in currency units, are 
the total amount of buying orders and selling orders, respectively. Note that 
for a single buying (or selling) order, the impact is given by $1/N$, i.e.
around $3 \%$ for a market with 30 participants (and less if the market 
involved more participants, as is reasonable). On the other hand, if
all the agents decide to buy (or sell) simultaneously, the ensuing impact factor would be $100\%$. 

With the introduction of market impact, subjects have 
to guess if the observed price fluctuations are due to 
``natural'' fluctuations, i.e. to the noise term they are warned about at the beginning of 
the game, or if they are due to the action of their fellow
subjects. This was meant to provide a potentially destabilizing channel, where 
mild sell-offs could spiral into panic and crashes. 

We are therefore interested in studying whether excess trading emerges in a market environment which clearly penalises trading, and whether synchronisation of trading activity, such as avalanches of selling orders triggered by panic, results in big market crashes. 

\subsection{Implementation}\label{subsec:implementation}
The experiment was programmed in Java using the PET software\footnote{PET software was developed by AITIA, Budapest, and is available at http://pet.aitia.ai.} and it was
conducted at the CREED laboratory at the University of Amsterdam in May
2014. In the beginning of the experiment each subject is randomly assigned a computer in the laboratory and physical barriers guarantee that there is no communication between subjects during the experiment. 
The experiment was conducted with 201 subjects, divided in 9 experimental markets with a median number of participants per market of 24 subjects.\footnote{The smallest market includes 15 subjects while the largest includes 29 subjects.} The experiment lasted about two hours and before starting the experiment subjects had to answer a final quiz to make sure they understood the rules of the game (see Appendix \ref{sec:instructions}).\footnote{For practical reasons we selected in advance exponential-distributed end-times to ensure that sessions would not stop too early.}

At the beginning of the experiment each subject is endowed with 100 francs.\footnote{In a pilot session we also implemented some markets in which subjects were endowed with 1 share worth 100 francs. This resulted in many subjects selling shares at the very beginning of the experiment, realising immediately net losses and subsequently engaging in trading activity trying to make at least some profits. While the difference arising from the different initial conditions is interesting in itself, we leave it for future work and focus on the case of initial endowment of cash.} Each period lasts 20 seconds, during which subjects have to decide whether they want to hold cash or shares in the next period.\footnote{In a pilot session we implemented a duration of 40 seconds per period. Decision times were, however, well below the threshold of 20 seconds, so we reduced the duration of each time step.} If subjects did not make any decision within the limit of 20 seconds, their current market position would simply carry over to the next period. If they have 
cash at period $t$, they can decide either to use it all to buy shares or to 
stay out of the market at period $t+1$. Conversely, if they have shares at 
period $t$, they have to choose between selling them all for cash or staying in 
the market at period $t+1$. Fractional orders are not allowed, therefore each 
player is either in the market or out of the market at all times.

In order to avoid possible ``demand effects'', i.e. 
trading volumes in the markets related to the fact that participation in the asset market is the only activity available for subjects \citep{lei2001nonspeculative}, we introduce a forecasting game in which subjects are asked to forecast the asset price and rewarded according to the precision of their forecast.\footnote{Moreover, we remark that practically speaking, the choice of being active or inactive in the market involved the same type of action, i.e. a click on the correspondent radio button (see Appendix \ref{sec:instructions} for details).}

The information visualised on subjects' screens include a chart displaying the asset price evolution together with a table reporting time series of asset prices, returns, individual holdings of cash and shares and price forecasts. Fig.~\ref{fig:screenshot} in Appendix \ref{app:extra_figs} shows a screenshot of the experiment.

In order to mitigate behaviour bias towards the end of the experiment, we implement an indefinite horizon (see \cite{crockett2013dynamic} for an example in artificial asset markets).\footnote{There is also a large experimental literature which implements a similar procedure to
study infinitely repeated games in the laboratory, beginning with \cite{Roth1978equilibrium}.}
In each period of the experiment there is a known constant continuation probability $1-p=0.99$. Moreover, each subject participates in two consecutive market sessions, after a short initial practice session to familiarise with the software, so that learning mechanisms can be studied.\footnote{In order to minimise changes in the market environments and facilitate subjects' learning we use the same noise realisations in the first and the second market sessions.}

When a market session is terminated, subjects' final wealth is defined as follows: if they are holding cash, their amounts of francs will determine their wealth; if they are holding shares, their wealth is defined as the market value of their shares, i.e. their amounts of shares times the market price at the end of the sequence. In fact, at the end of the market sequence shares are liquidated at the market price without any price impact.\footnote{In this way, for a large number of periods, the average final wealth of a risk-neutral population would be approximately the same if agents were given shares instead of cash in the beginning of the session.} At the end of the experiment subjects' \textit{net} market earnings are computed as their end-of-sequence balance minus their initial endowment. At the end of the experiment, each subject 
rolls a dice to determine which of the two market sessions will be used to calculate 
his take-home profits (which also include the earnings from the forecasting game). The exchange rate is 100 francs to EUR 25. 

Finally, we asked subjects to participate in
a second experiment involving the \cite{holt2002risk} paired lottery choice instrument.
This second experiment occurred after the asset market experiment had concluded and
was not announced in advance, to minimize any influence on decisions in the asset market
experiment.

The complete experimental instructions can be found in Appendix \ref{sec:instructions}.

\section{Theoretical Benchmark}\label{sec:rational_bench}
In this section we devote attention to the theoretical rational benchmark for this experiment and derive the equilibrium of the game under the assumption of full rationality and common knowledge of rationality. It is worth remarking at this point that this experiment is a {\it de facto} risk-free opportunity, in the sense that the subjects are paid at the end of the experiment if their net profit is positive but do not owe any amount if their net profit ends up negative.

\subsection{Risk neutrality}\label{subsec:risk_neut}

Let the wealth of the agent be $w_{t}$ at time $t$. If we assume that the session ends at $t=t_F$, 
fully rational agents have two possible strategies. The first one is to stay out of the market for $t > 1$ 
and hold cash until $t=t_F$ which yields an expected final wealth $\mathbb{E}\left(w_{t_{F}}\right)=w_{0}$.

The second strategy consists in being fully invested in the market 
at $t=1$ and hold the shares until $t=t_F$. In fact, in every period holding shares is an equilibrium of the stage game since the strategy ``Hold'' is the best response to any number of players $n$ that an agent might expect to sell.\footnote{In fact, given the set of actions $\mathcal{A}_t = \{\mbox{Hold,Sell}\}$ available to the agent holding shares, the action ``Hold'' vs. ``Sell'' would result in a price $p^H_{t+1}=p_t\cdot \exp(m-(n-1)/N + s\eta_t)$ vs. $p^S_{t+1}=p_t\cdot \exp(m-n/N + s\eta_t)$, with $p^H_{t+1}>p^S_{t+1}$.} This strategy yields $\mathbb{E}\left(w_{t_{F}}\right) \approx w_{0} \cdot (1+m+s^2/2)^{t_F}$ 
regardless of the initial condition. The average outcome of the first strategy 
is at best zero profit (remember that subjects only keep net profits), while the second strategy provides large expected
profits when the experiment session lasts for a long time.

Therefore the rational benchmark, with a risk neutral population, is to get in the market at $t=1$ and hold the shares until the end of the experiment.\footnote{We remark however that, due to the indefinite horizon nature of the game, we can not rule out the presence of other equilibria on the basis of the Folk theorem. Nevertheless, coordination on such equilibria would require an extremely demanding coordination device for subjects. Given the laboratory environment in which the game is played (not to mention the presence of the noise in the price process), coordination on these equilibria can be ruled out as a concrete possibility.} The longer the duration of the experiment, the larger the expected profits.
 
\subsection{Myopic risk aversion}\label{subsec:risk_aver}

Empirical evidence suggests that risk attitude affects trading behaviour. In the following we consider the case of risk averse traders using the utility function proposed by \cite{holt2002risk} 
\begin{equation}\label{eq:utility}
	U(w)=\frac{1-\exp\left(-\alpha w^{1-r}\right)}{\alpha},
\end{equation}
where $\alpha$ and $r$ are positive parameters.\footnote{See also \cite{harrison2007naturally} and \cite{harrison2008risk} for an overview on risk aversion in the laboratory.} Based on the fact that the utility function in Eq.~\eqref{eq:utility} is 
concave, one could be tempted to consider that for high
enough volatility $s$, myopic rational traders, i.e. traders who only look one time step ahead when making their
decisions, would choose to step out of the market as soon as their wealth reached a certain level. However,
this is not the case because if they decided to sell, the price of the asset would not only be negatively affected by
their action through the impact factor $I_t$, but also be impacted by the noise $s\eta_t$. This is due to the fact that transactions ordered at time $t$ are executed at the price realized in $t+1$, meaning that agents deciding to step out of the market at time $t$ are still subject to price volatility. Consequently, it is clear that, for a
given time step, it is always a better option to stay in the market and be affected by the noise factor alone, instead of selling and being affected by both the noise factor and the negative impact factor.\footnote{In fact for any number of players $n$ that an agent might expect to step out of the market, the actions ``Hold'' vs. ``Sell'' available at time $t$ to an agent holding shares would result in a price $p^H_{t+1}=p_t\cdot \exp(m-(n-1)/N + s\eta_t)$ vs. $p^S_{t+1}=p_t\cdot \exp(m-n/N + s\eta_t)$, with $p^H_{t+1}>p^S_{t+1}$, meaning that $E[U(\mbox{Hold})]=E[U(p^H_{t+1})]>E[U(\mbox{Sell})]=E[U(p^S_{t+1})]$. The only case in which an agent can receive a certain amount of wealth with certainty is at the beginning of the experiment by deciding to keep the initial endowment and staying out of the market until the end of the experiment. This strategy would however result in zero profits since subjects can only keep net profits.} 
%
%
%
%

%
%one can consider scenarios in which, for high enough volatility $s$, 
%myopic rational traders with heterogeneous risk preferences, i.e. traders who only look one time step ahead when making their
%decisions, would choose to step out of the market as soon as 
%their wealth reaches a certain level.\footnote{In the case of homogeneous risk preferences, rational agents would anticipate that if they all decided to sell, the price of the asset would be negatively 
%affected by the impact factor $I_t$ on top of the unknown noise term $s\eta_t$. Consequently, it would be a better option to stay in the market and be affected by the noise 
%factor alone, instead of selling and being affected by both the noise factor 
%and the negative impact factor.} This would lead to initially increasing price impacted by noise only, followed eventually by downwards price corrections triggered by agents stepping out of the market according to their degree of risk aversion. 

Using Eq.~\eqref{eq:utility} we can also study a few cases of bounded rationality within myopic optimisation. In particular we compare, for different levels of wealth, the expected utility given by Eq.~\eqref{eq:utility} in the following anticipated scenarios: all players hold to their shares, taking into account the volatility of the noise term ($I=0$); all players sell their shares, taking into account the volatility of the noise term ($I=-1$); all players sell their shares, without taking into account the volatility of the noise term ($I=-1$ no noise); one player only sells his shares, without taking into account the volatility of the noise term ($I=-0.01$ no noise). 

These scenarios are compared in Fig.~\ref{fig:utility} below.\footnote{Fig.~\ref{fig:utility} is obtained using the parameter values estimated in \cite{holt2002risk} for the power-expo utility function, namely $\alpha = 0.029$ and $r = 0.269$. The parameter estimates obtained using the lottery choices of our subjects' sample, i.e.~$\alpha = 0.106$ and $r = 0.345$, are -- remarkably -- close to the values estimated by \cite{holt2002risk}, see Appendix \ref{app:risk_est}.}
The magnitude of the noise $s$ defines most of the differences.
\begin{figure}[!htb]
	\subfloat[RMS $s=0.1$\label{fig:utility_s0.1}]{%
		\includegraphics[scale=0.3]{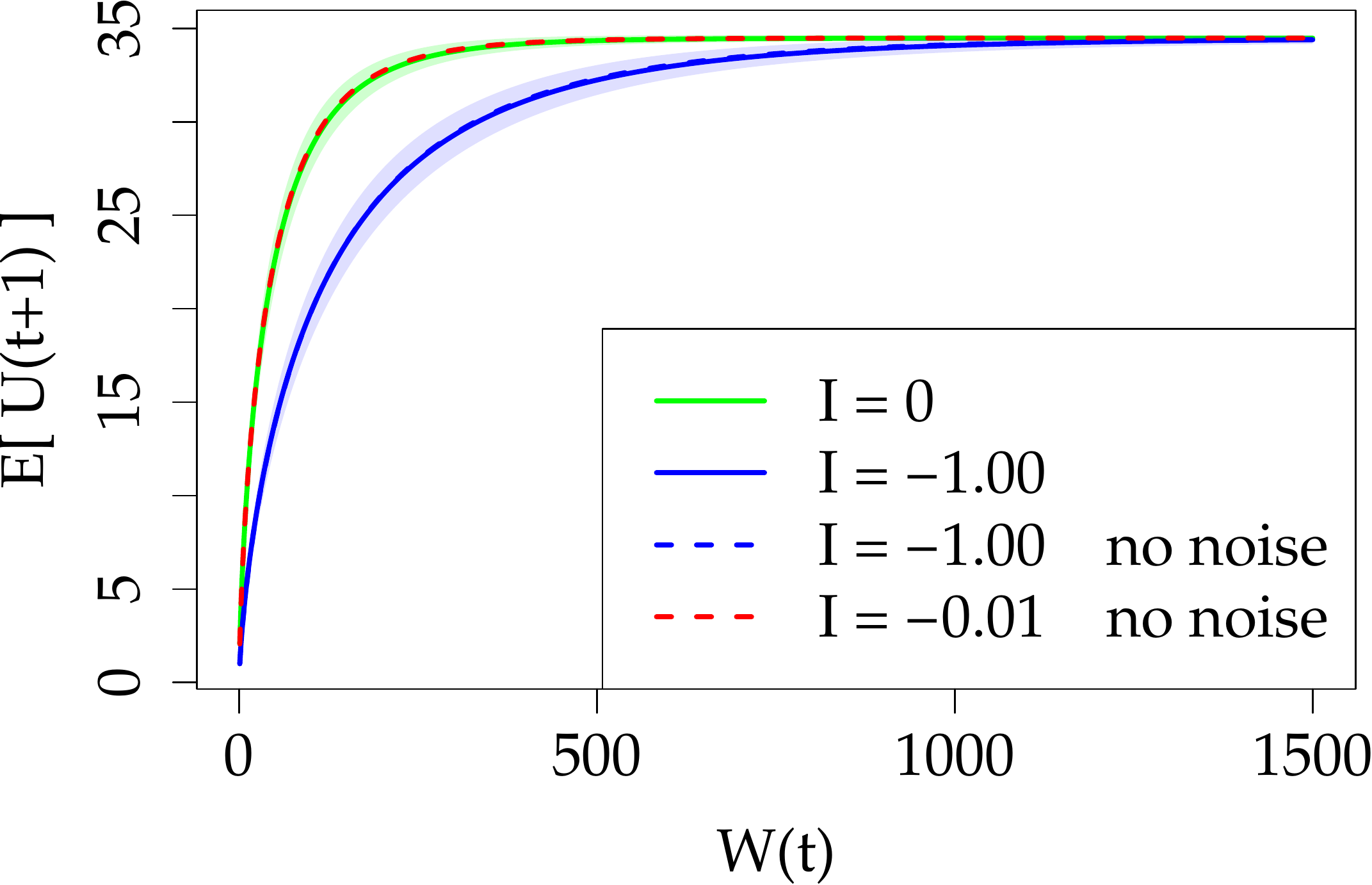}%
		}\hfill
	\subfloat[RMS $s=1.0$\label{fig:utility_s1.0}]{%
		\includegraphics[scale=0.3]{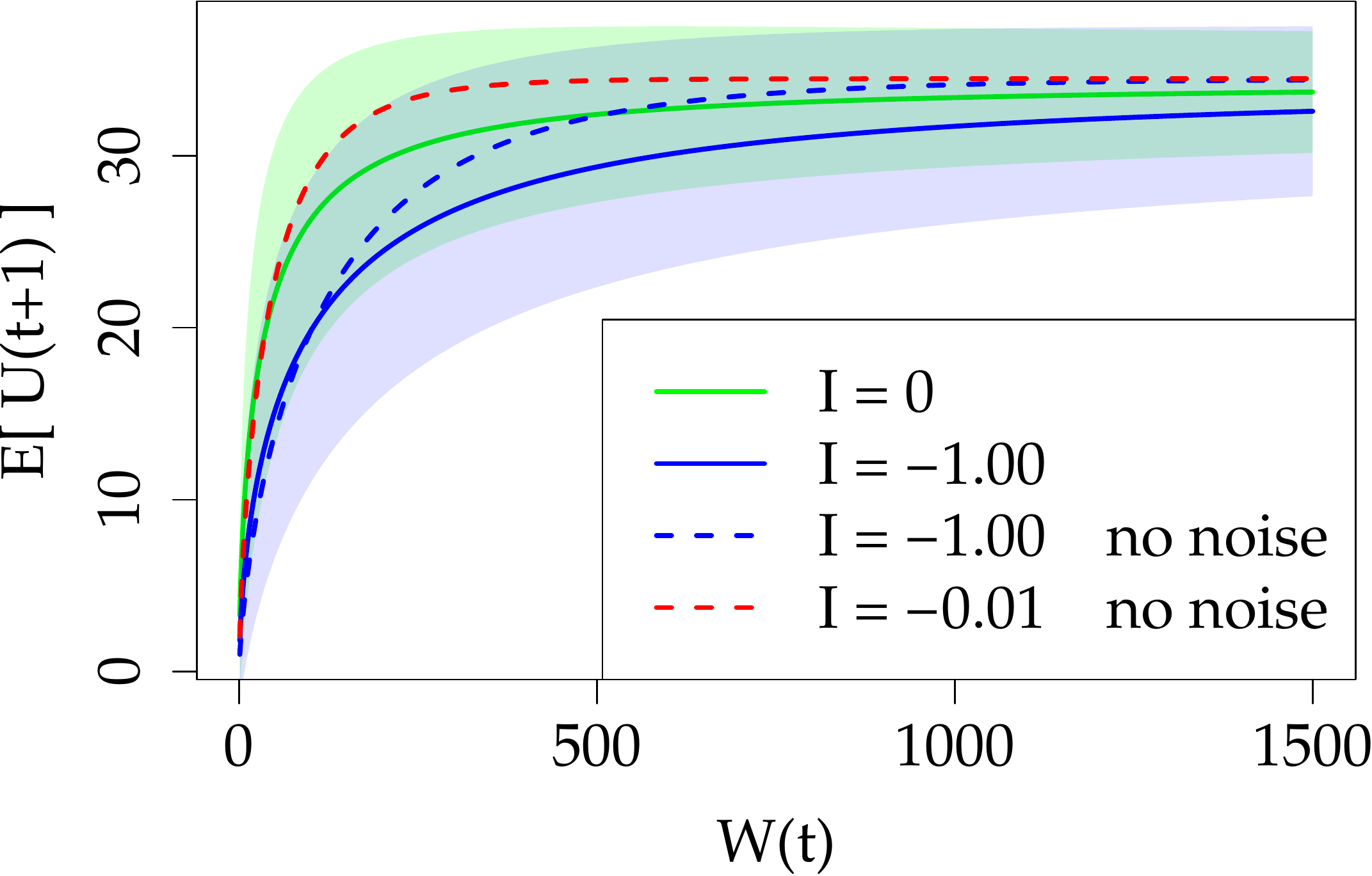}%
	}
	\caption[Content]{Expected utility in Eq.~\eqref{eq:utility} for 
different scenarios with bounded rationality, volatility $s=0.1$ (left) and $s=1.0$ (right) 
and varying level of net wealth $W(t)$. The coloured regions
correspond to 1-$\sigma$ around the average utility.}
	\label{fig:utility}
\end{figure}
Fig.~\ref{fig:utility} gives insight on the possible behaviour of risk-averse players. Comparing the solid lines we confirm the intuition presented above: these solid lines never intersect and it is always better for all agents to stay in the market (green solid line) than to step out collectively (blue solid line). However, when we consider bounded rationality and myopic risk-averse traders, the conclusion may be different, in particular when the random fluctuations of the market, which impact the price at each time step, are not taken into account. Consider for example the utilities, computed without taking into account the random fluctuations, for the cases in which only one agent sells (red dashed line) and everyone sells (blue dashed line), and compare them with the case in which everyone stays in the market (green solid line). In these scenarios, there will be a value of wealth $W(t)$ above which, in the 
mind of these boundedly rational agents, it pays off to sell whatever they are holding. 
This point depends on the agent thinking either that he will be the only one selling 
or that everyone will, as well as on the magnitude of the random fluctuations 
or noise $s$. These values are critical thresholds which we represent as a 
function of $s$ in Fig.~\ref{fig:utility_intersect} through numerical 
simulations. \begin{figure}[!htpb]
\centering
	\includegraphics[scale=0.3]{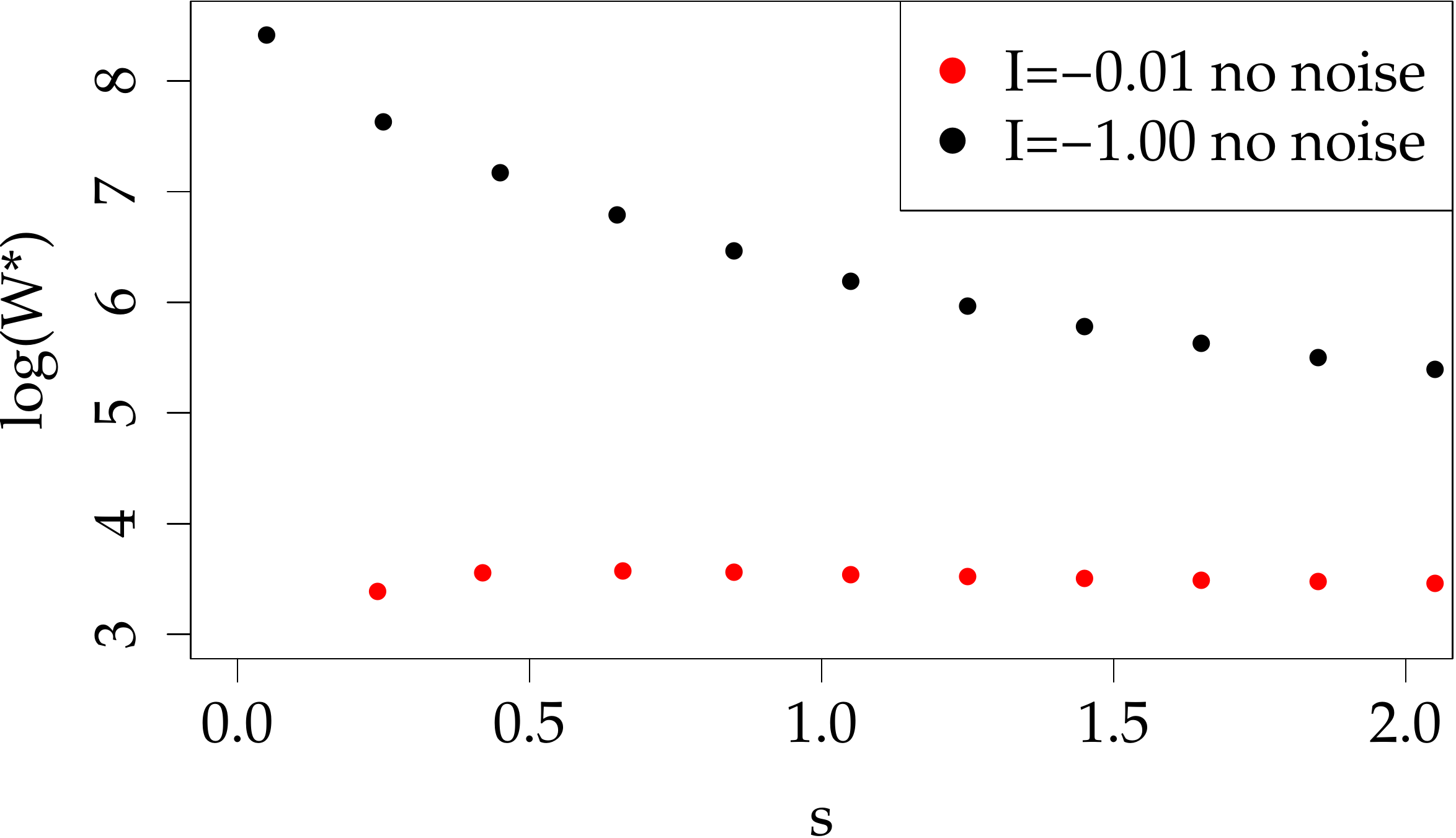}%
	\caption[Content]{Critical threshold values $W^*$, as a function of $s$, beyond which the 
single-player-out (in red) and all-out strategies (in black) are triggered (corresponding, respectively, to the intersection point of 
	the green/red dotted lines and green/blue dotted lines in Fig.~\ref{fig:utility}).}
	\label{fig:utility_intersect}
\end{figure}
As we would expect from the concavity of the utility function in 
Eq.~\eqref{eq:utility}, the value $W^*$ beyond which boundedly rational agents of this 
sort would sell decreases as a function of $s$ when they consider the 
possibility of everyone selling at the same time.

In the presence of myopic boundedly rational agents of the sort considered in the section, heterogeneity in risk preferences would lead to threshold strategies in which traders, when wealth increases beyond a critical value $W^*$, get nervous and get out of the market. This scenario does not, however, account for the variability amongst agents, which most certainly impacts 
observed events in real markets and, as we shall see below, the trading activity observed in our experiment. 

%In summary, we showed that heterogeneous risk preferences or bounded rationality lead to threshold strategies in which traders, when wealth increases beyond a critical value $W^*$, get nervous and get out of the market. 
%This does not, however, account for the variability amongst agents, which most certainly impacts 
%observed events in real markets and, as we shall see below, the trading activity observed in our experiment. 

\section{Experimental Results}\label{sec:results}
\subsection{Trading activity}\label{ssec:global}
In this section we report on the trading activity observed in the experiment. Subsection \ref{sssec:wealth} describes results on the volume of trading activity in relation to subjects' wealth. Subsection \ref{sssec:collective} studies synchronization in subjects' trading activity, while Subsection \ref{sssec:clusters} investigates the presence of common patterns in subjects' trading behaviour.  

\subsubsection{Excess trading and wealth}\label{sssec:wealth}
\begin{figure}[!htpb]
\begin{center}
	\subfloat[First sessions: quantities averaged over 198 subjects (18 for the last point). The correlation between 
realized price log returns and price log returns in the absence of trading is 
$0.85$.\label{fig:c2_pt}]{%
		\includegraphics[scale=0.3]{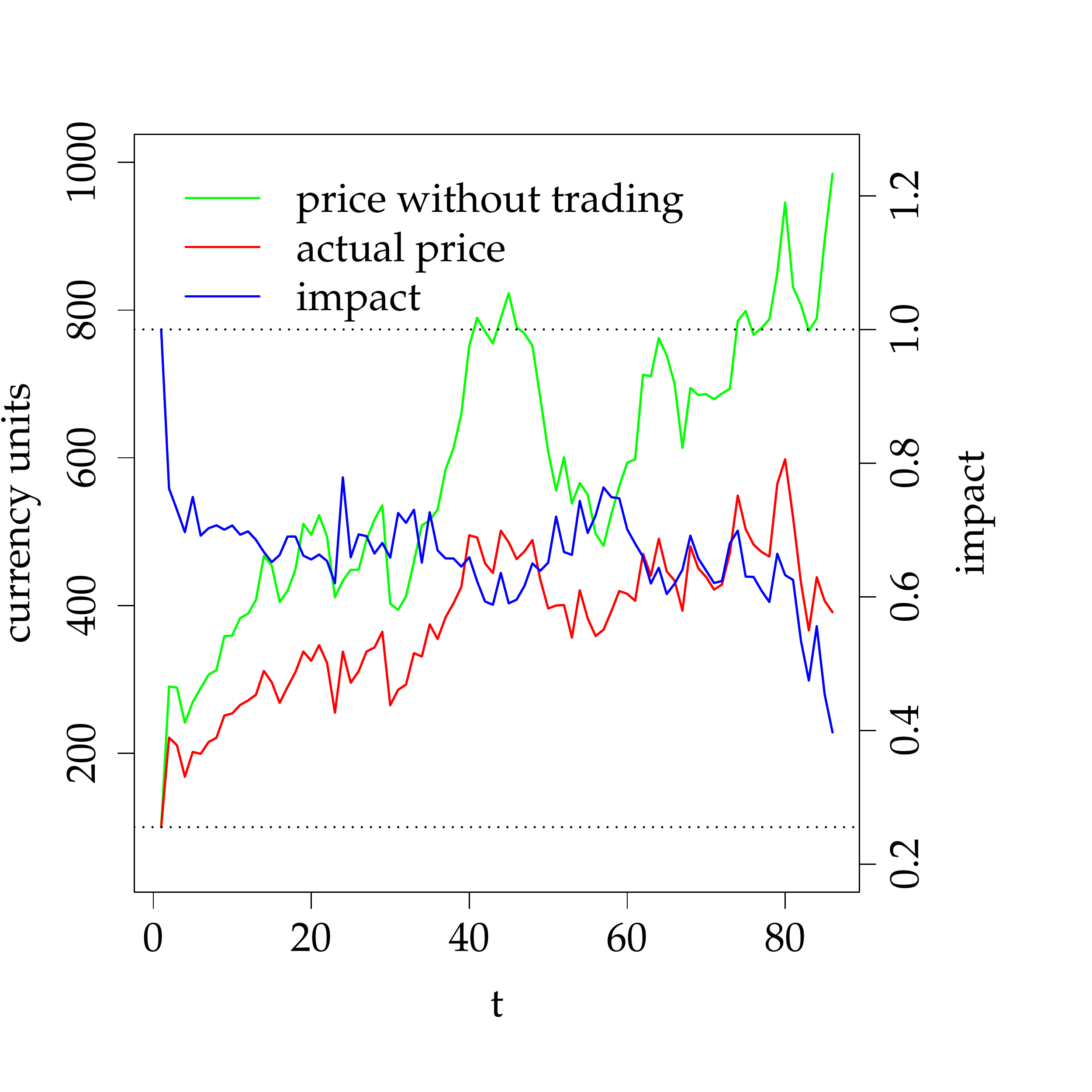}%
		}\hfill
	\subfloat[Second sessions; quantities averaged over 201 subjects (23 for the last point). The correlation between 
realized price log returns and price log returns in the absence of trading is 
$0.89$.\label{fig:c3_pt}]{%
		\includegraphics[scale=0.3]{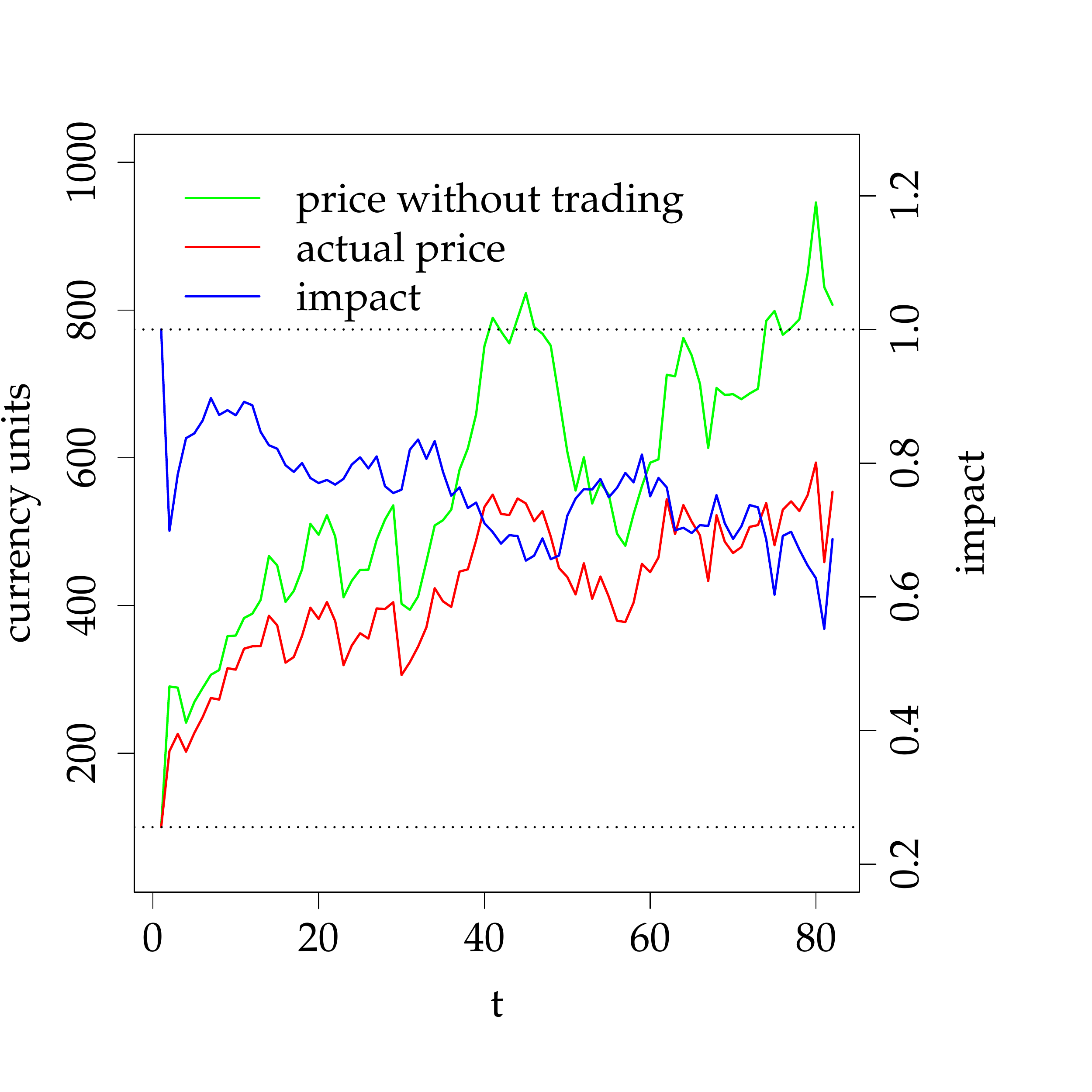}%
	}
	
\end{center}
	\caption[Content]{Average price time series for first and 
second market sessions.}
	\label{fig:c2c3_pt}
\end{figure}

It is clear that if all subjects adopted the strategy of buying and holding the shares until the end of the game, the price impact would be $I(t)\equiv0$ and the realized price at the end of the market sequence would lead to a most probable $640\%$ increase in wealth. What we observe instead is that subjects keep on trading in and out of the market. The trading activity is in fact so high in the first market sessions that they barely break even, earning a meagre $0.75\%$ on average. Remarkably, all groups learn to some extent and trade much less in the second market sessions, leading to a much higher average earning of $92\%$.\footnote{In practice the average payouts were higher since participants could not incur losses, therefore negative contributions did not play a role in actual average payouts.} These results are somewhat in line with findings in the literature on experimental asset markets showing that sufficient experience with an asset in certain environments eliminates mispricing and the emergence of bubbles \citep{smith1988bubbles,king1991private,vanboening1993price,dufwenberg2005bubbles,haruvy2007traders,lei2009market}.\footnote{On the contrary, \cite{hussam2008thar} and \cite{Xie2012bubbles} find that bubbles
can be rekindled or sustained when the market experiences a shock from increased liquidity, dividend
uncertainty and reshuffling, or from the admission of new subjects.} In fact we observe, as a result of learning, consistently different behaviour of subjects in the second market sequences when compared to their behaviour in the first sequences. The Welch two-sample t-test applied to the final wealth and 
average activity rate of each subject in first and second runs statistically 
confirms this difference ($\text{p-value}\approx2^{-16}$). Therefore in the following analysis we merge all first market sessions into one dataset and all second sessions into another dataset. These aggregated data sets lead to the price time series illustrated in 
Fig.~\ref{fig:c2c3_pt}.
We remark that the durations of the 
sessions were all different because of the indefinite time horizon. 
Thus, the number of data 
points used in the averaging process is not the same for each time $t$ but a 
decreasing step function of $t$ (this is clearly visualised in Fig.~\ref{fig:positions} below).\footnote{Moreover, during the first experimental market sessions we experienced network problems with three computers. Hence the number of subjects included in the pool for the first runs is 198 instead of 201.} 

The lines in green represent the ``bare'' price time series, i.e.~the price evolution that would have occurred in the market if all agents had played the buy-and-hold strategy. A comparison with the red lines, depicting the realized prices, immediately reveals that trading significantly weakens the upwards price trend via the impact factor (blue lines): the average slope (i.e. price trend) is divided by a factor of approximately $2$ in the first sessions and $1.5$ in the second 
sessions. The realized prices in Fig. \ref{fig:c2c3_pt} are proxies for the 
maximum earnings a subject would achieve if he used the buy-and-hold strategy 
within his group. However, the realized price log-returns remain highly 
correlated
with the ``bare'' price time series: the correlation coefficient is $0.85$ in 
first market sessions and $0.89$ in second market sessions.
The higher values of the slope (i.e. price trend) and of the correlation coefficient for the 
second sessions are due to a lower trading activity.

Fig.~\ref{fig:positions} displays the positions of the traders -- in the market (green) or out of the market (red) -- throughout the experimental market sequences. 
\begin{figure}[!htpb]
\begin{center}
	\subfloat[First sessions: 198 subjects.\label{fig:c2_p}]{
		\includegraphics[scale=0.3]{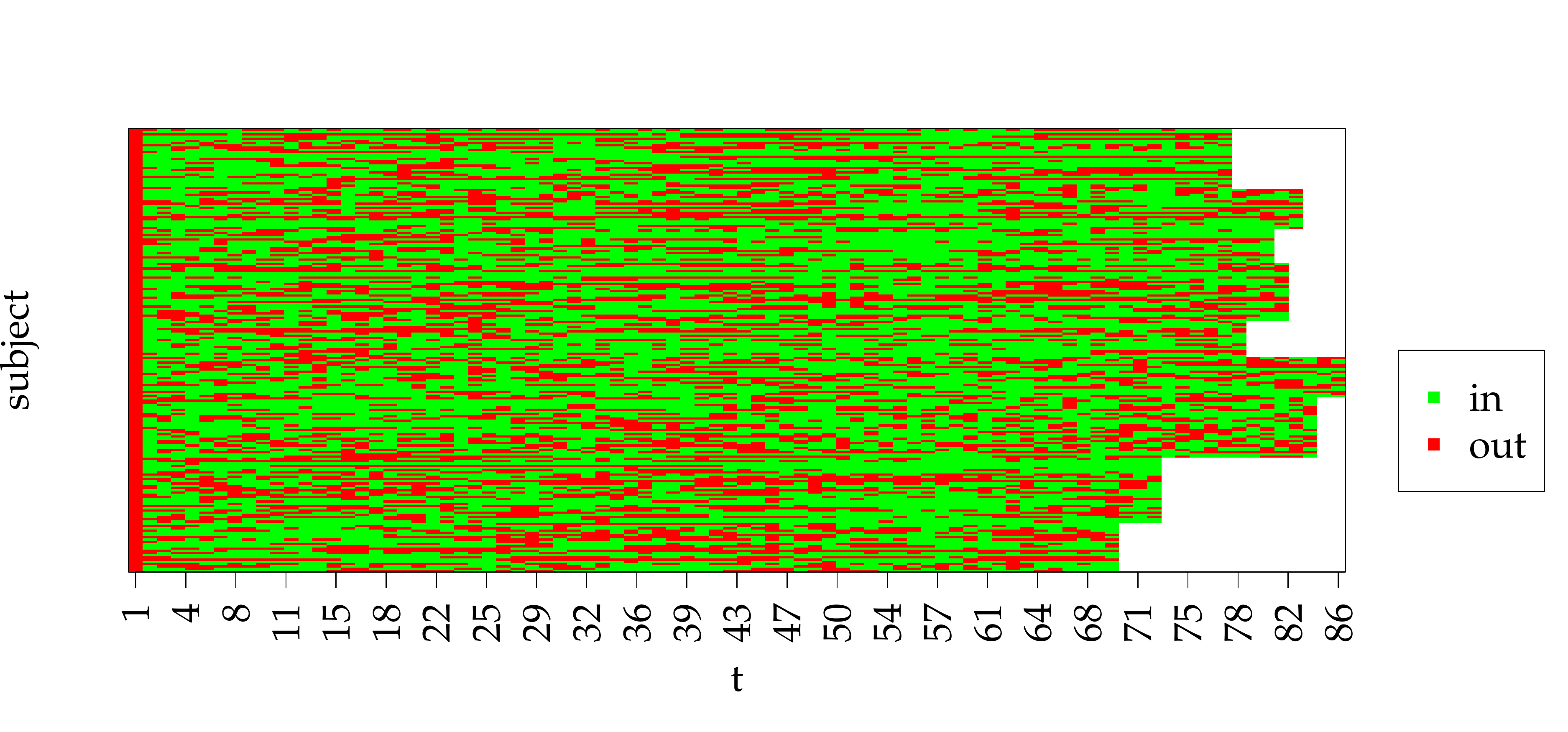}
	}
	\\
	\vspace{-0.5cm}
	\subfloat[Second sessions: 201 subjects.\label{fig:c3_p}]{
		\includegraphics[scale=0.3]{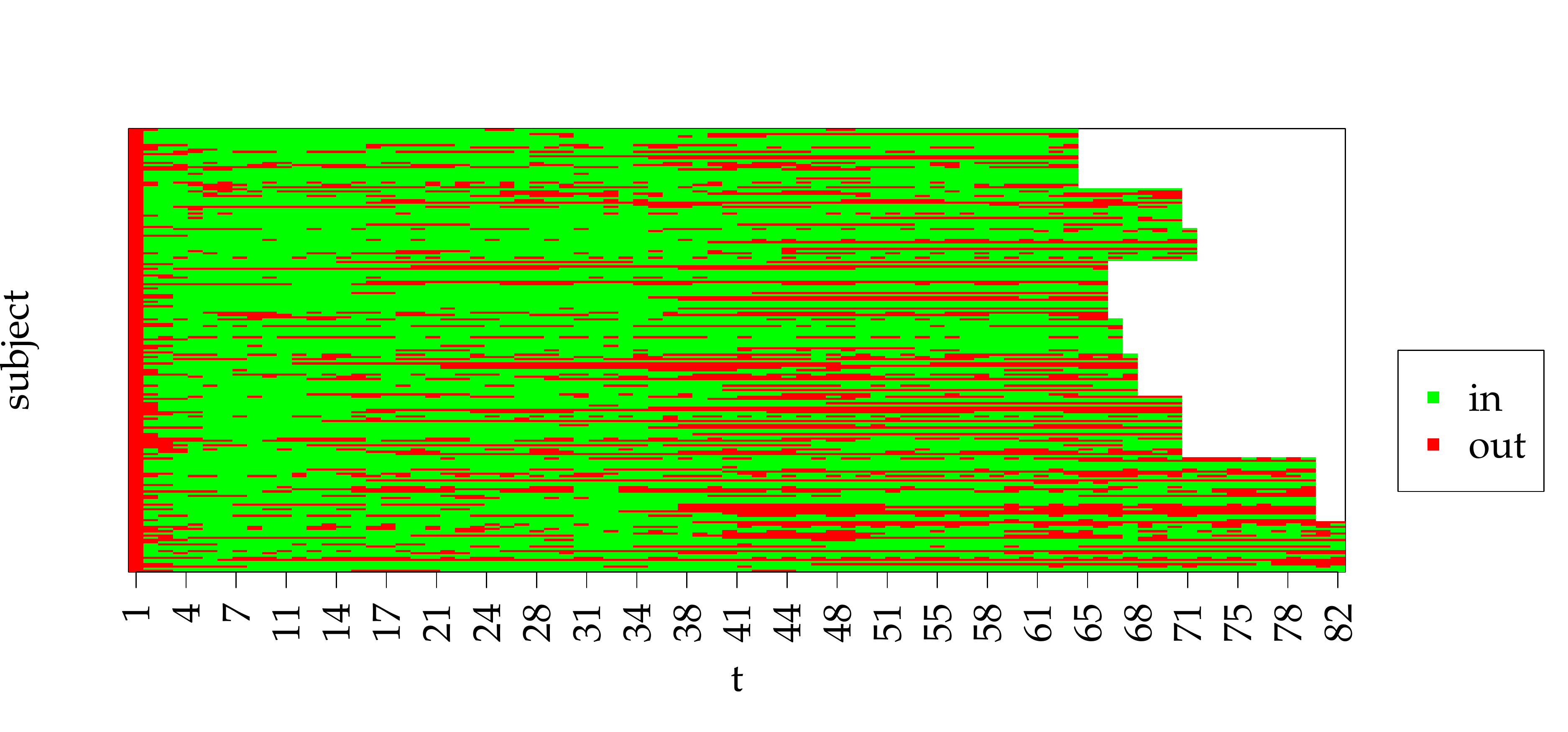}
	}
	\end{center}
	\caption[Content]{Traders' positions -- in the market (green) or out of the 
market (red) -- in first sessions (above) and in second sessions (below).}
	\label{fig:positions}
\end{figure}
Notice
that not all the sessions lasted the same number of periods, hence the white space 
in both Figs.~\ref{fig:c2_p} and Fig.~\ref{fig:c3_p} for large $t$. In fact, for each case, only one session -- the longest -- 
lasted until the maximum time $t$ displayed, $t=86$ for first sessions (Fig.~\ref{fig:c2_p})
and $t=82$ for second sessions (Fig.~\ref{fig:c3_p}).

Subjects' positions -- in (holding shares) or out of the market (holding cash) -- are mostly intermittent, which implies excessive trading. 
However, comparing Figs.~\ref{fig:c2_p} and \ref{fig:c3_p}, we observe that when the same subjects play 
for a second time, some of them actually learn the optimal 
strategy, which translates into ``green corridors'' in Fig.~\ref{fig:c3_p}.

\begin{figure}[!t]
\centering
	\subfloat[First sessions: 198 subjects and 69-85 periods.  The average 
final wealth was $100.75$ units of currency, the average activity rate $29\%$ 
and the correlation between the two $-0.62\%$.\label{fig:c2_a}]{%
		\includegraphics[scale=0.3]{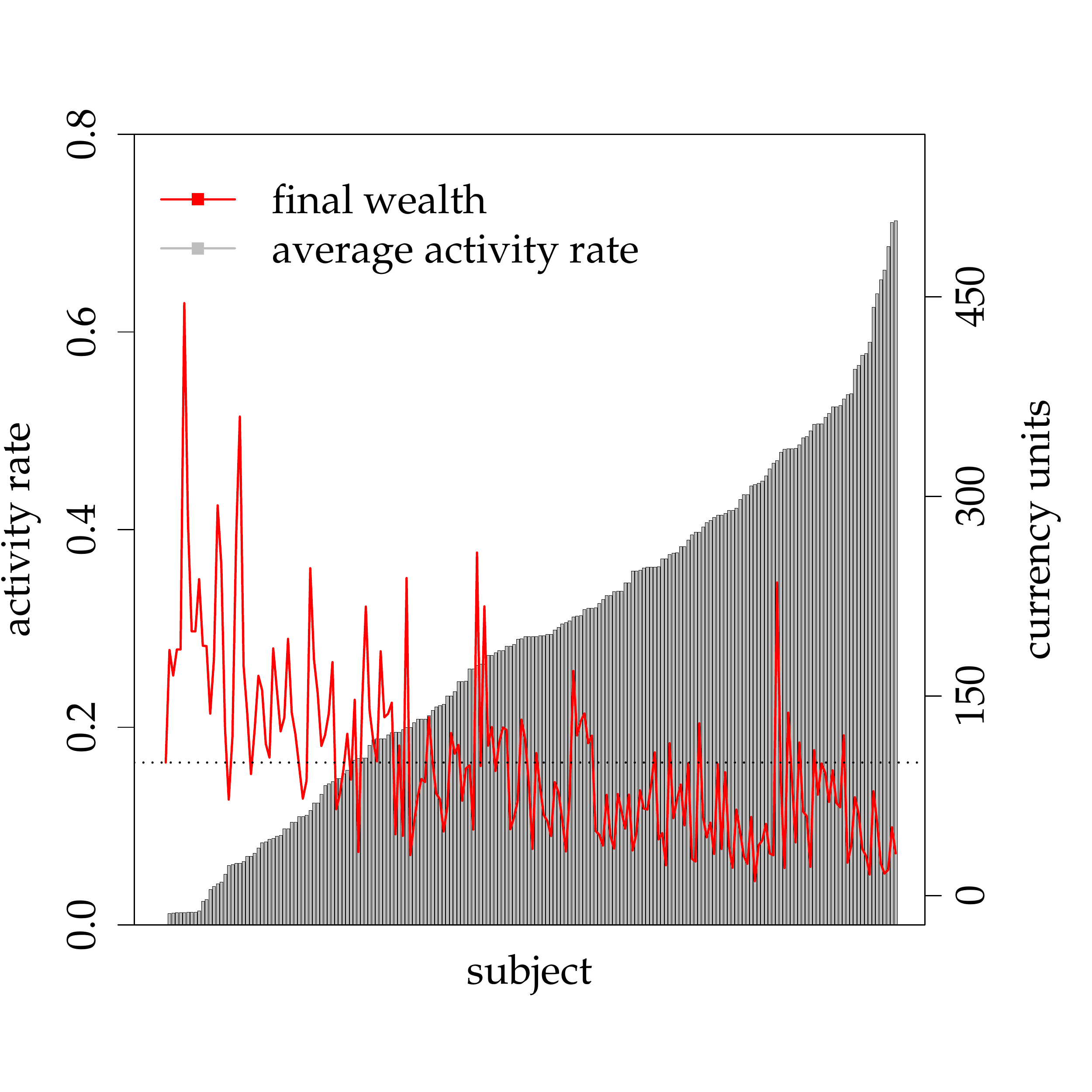}%
		}\hfill
	\subfloat[Second sessions: 201 subjects and 63-81 periods. The average 
final wealth was $191.97$ units of currency, the average activity rate $12\%$ 
and the correlation between the two $-0.73\%$.\label{fig:c3_a}]{%
		\includegraphics[scale=0.3]{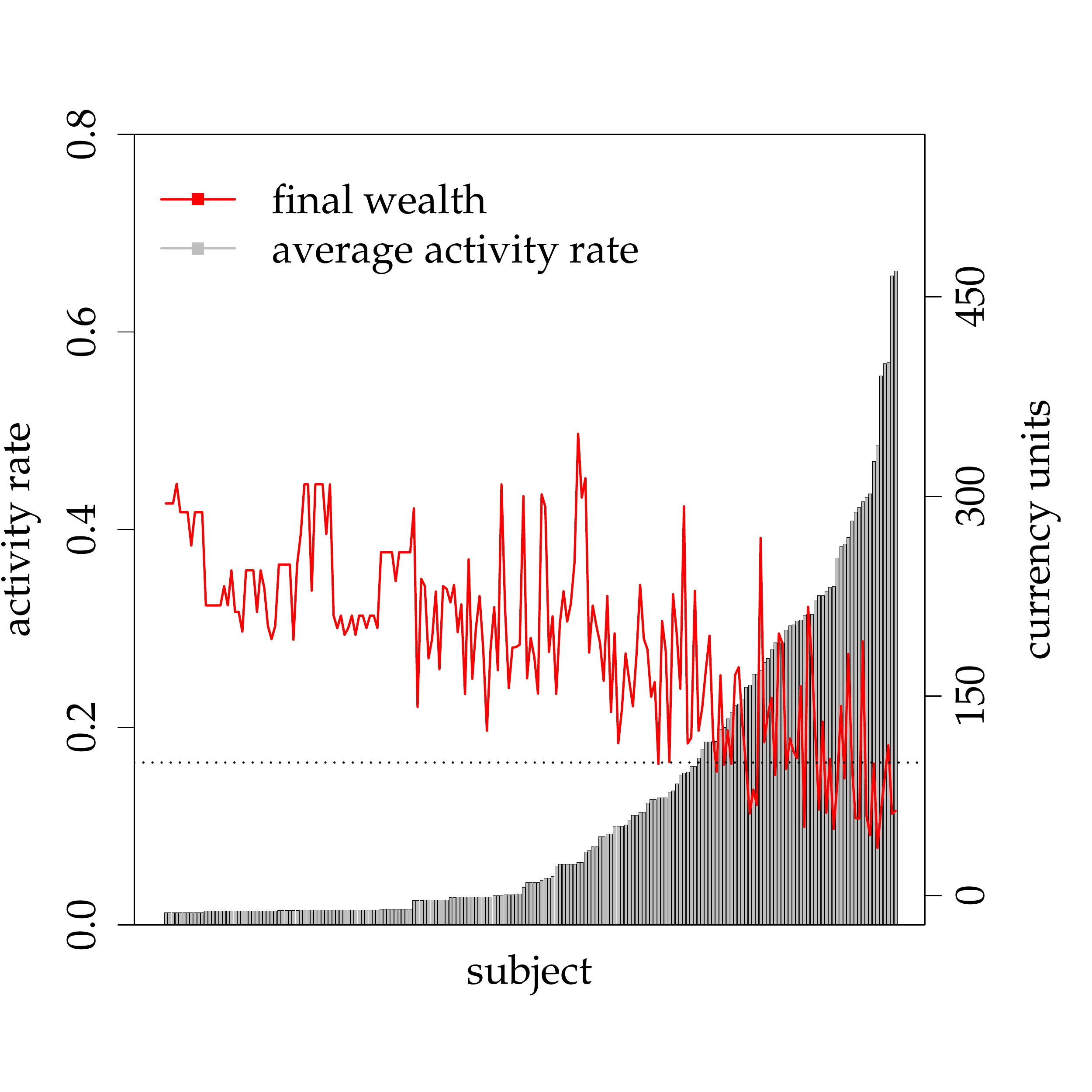}%
	}
	\caption[Content]{Activity rate and final wealth in first sessions (left) 
and in second sessions (right).}
	\label{fig:activity}
\end{figure}

Fig.~\ref{fig:activity} shows that the distribution of average trading activity changes when the same 
set of people play the game for the second time, and relates it to agents' final wealth. 
The number of people who keep trading 
to a minimum increases significantly in second sessions, where only a few 
outliers keep trading activity above $40\%$, i.e. they changed their market positions
in more than $40\%$ of the periods. In both cases, the 
final wealth of the agents is strongly anti-correlated with average trading 
activity, which is expected since trading is costly. 
In fact, if a trader decides to buy shares at period $t$ and to sell them at period $t+1$, 
he will, on average, end up with less cash than he started because of his own contribution to price impact. 

\begin{figure}[]
\centering
	\subfloat[First sessions, 198-18 subjects.\label{fig:c2_wt}]{%
		\includegraphics[scale=0.3]{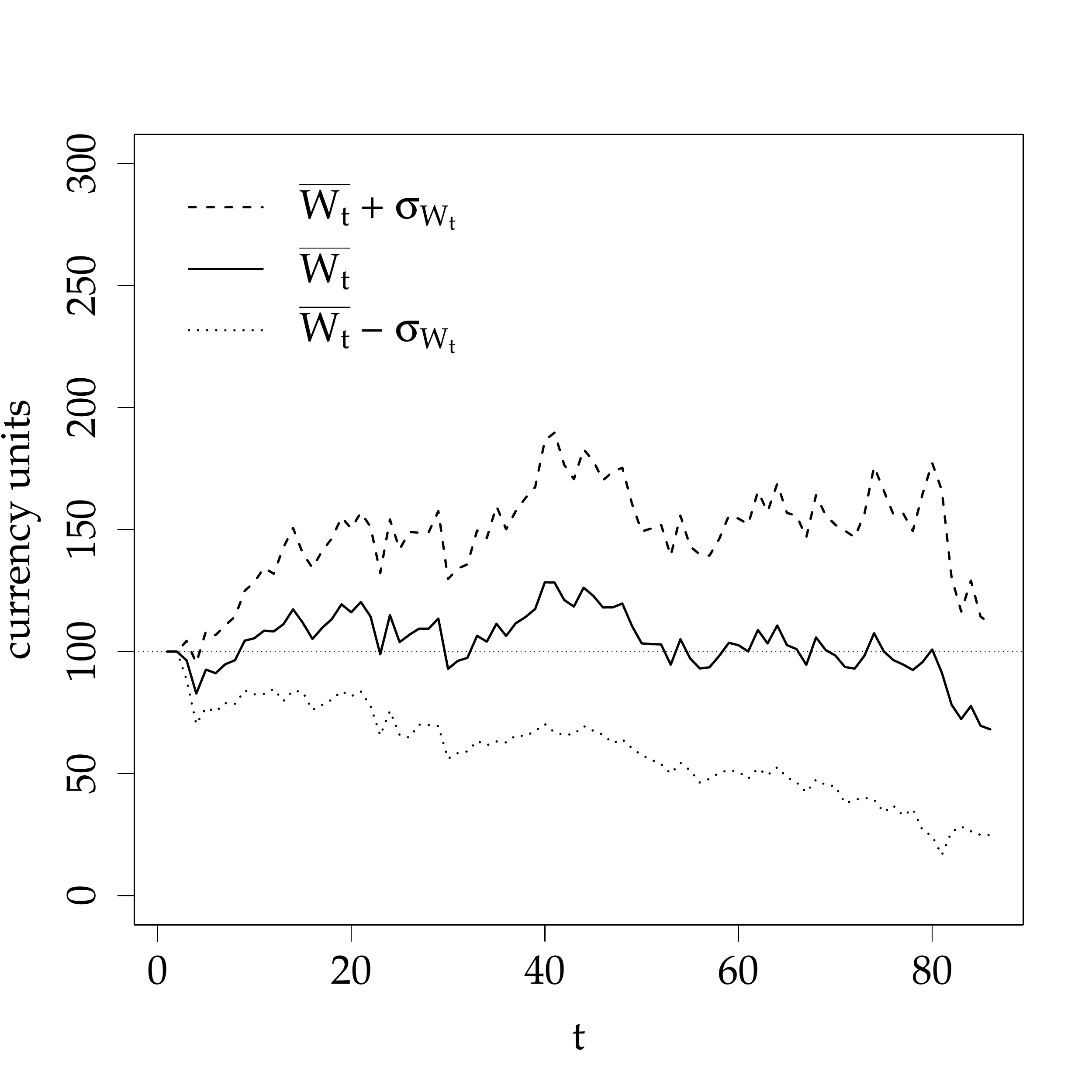}%
		}\hfill
	\subfloat[Second sessions, 201-23 subjects.\label{fig:c3_wt}]{%
		\includegraphics[scale=0.3]{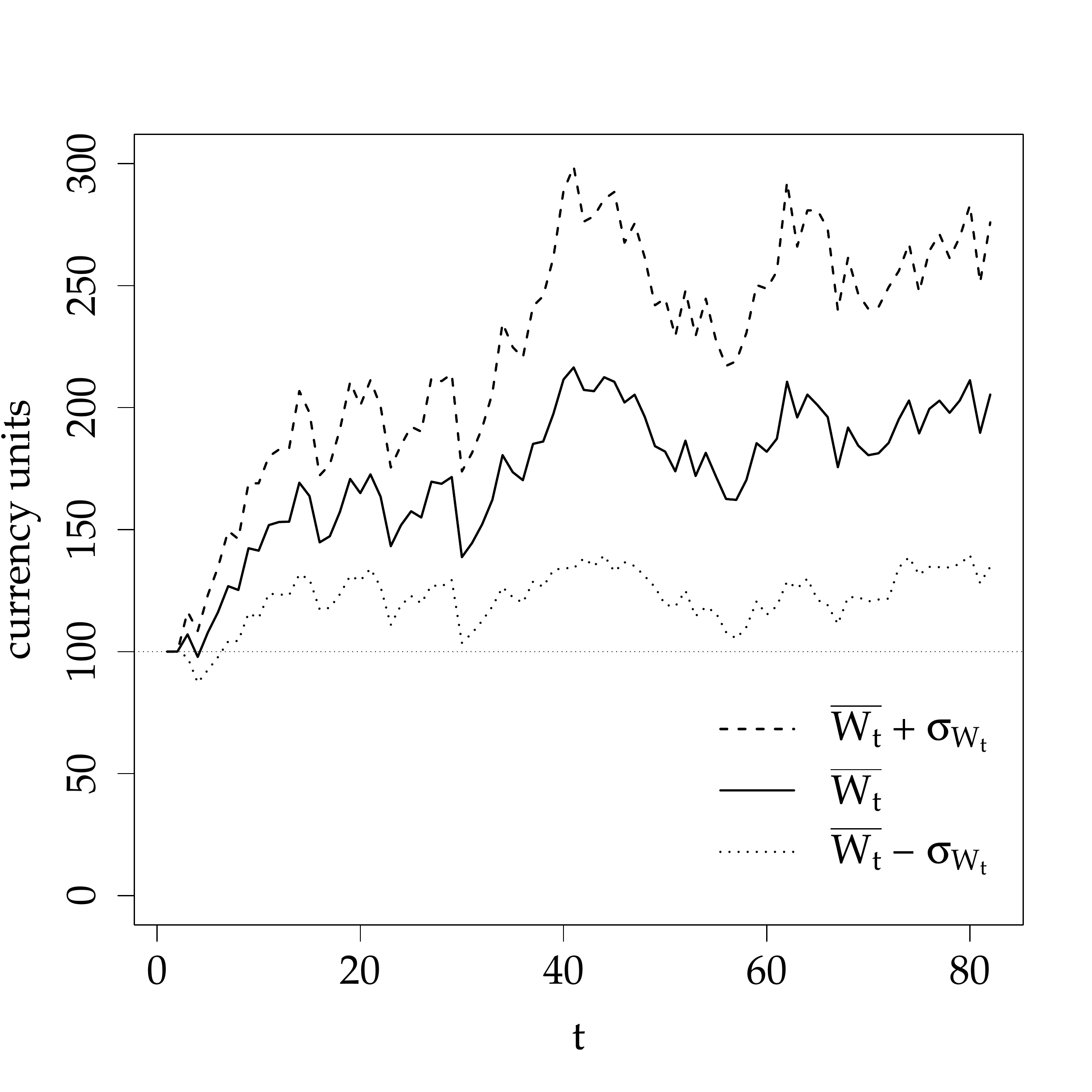}%
	}
	\caption[Content]{Average wealth for first sessions (left) and second sessions (right).}
	\label{fig:c2c3_wt}
\end{figure}

\begin{figure}[!htpb]
\centering
	\subfloat[First sessions, 198-18 subjects.\label{fig:c2_wd}]{%
		\includegraphics[scale=0.3]{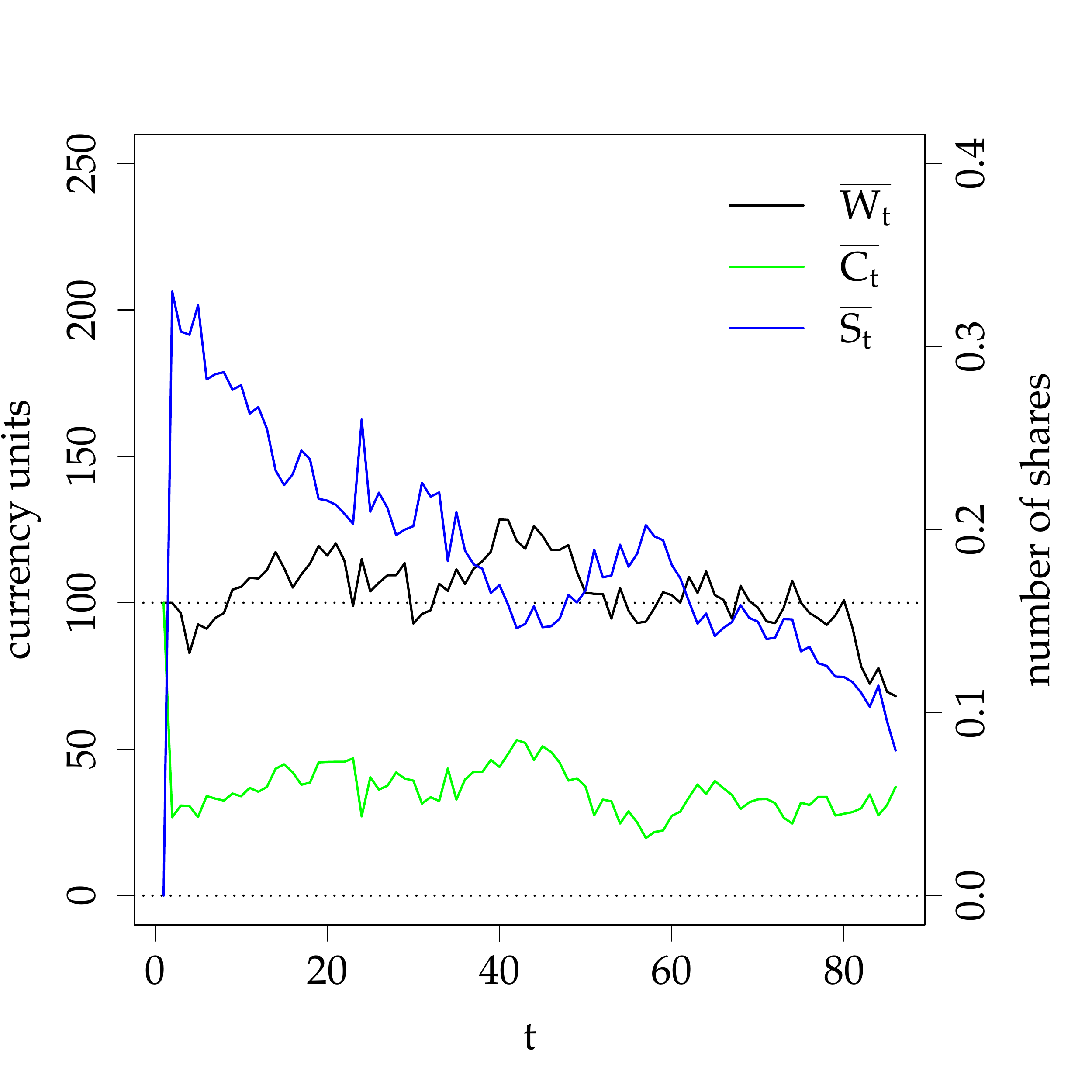}%
		}\hfill
	\subfloat[Second sessions, 201-23 subjects.\label{fig:c3_wd}]{%
		\includegraphics[scale=0.3]{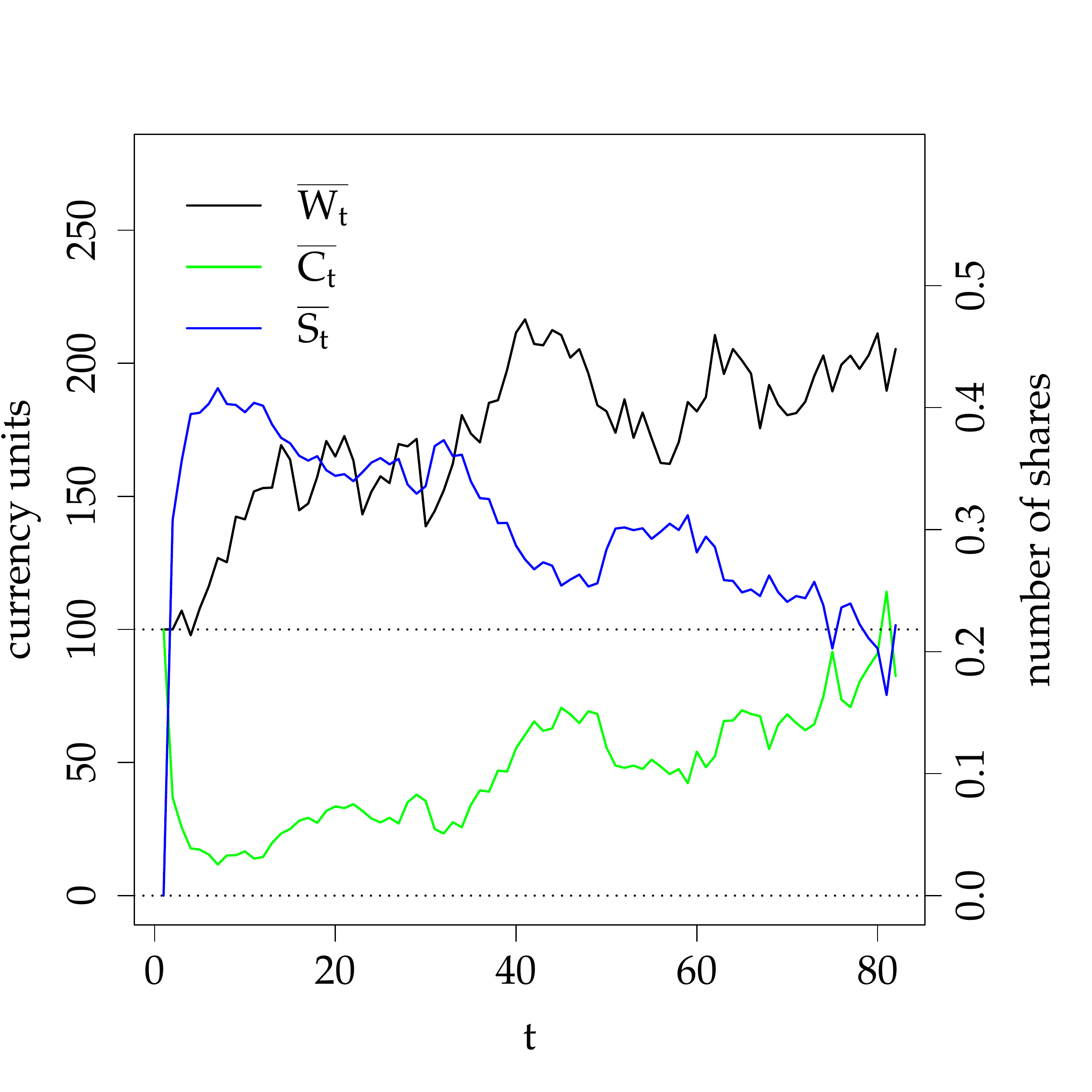}%
	}
	\caption[Content]{Average wealth $\overline{W_t}$ (black), average cash 
$\overline{C_t}$ (green) and average number of shares $\overline{S_t}$ (blue) 
in first sessions (left) and in second sessions (right).} 
	\label{fig:c2c3_wd}
\end{figure}

The average wealth of the subjects in first and second sessions is shown in 
Fig.~\ref{fig:c2c3_wt} as a function of time, while Fig.~\ref{fig:c2c3_wd} shows the average components of wealth over time. 

Fig.~\ref{fig:c2c3_wt} shows once more that subjects fared much better in second sessions, in which their average final wealth was approximately twice their initial endowment, than in first sessions, in which the vast majority did not even break even. In what concerns the average components of wealth over time, there are also 
differences between first sessions (Fig.~\ref{fig:c2_wd}) and second sessions 
(Fig. \ref{fig:c3_wd}). In first sessions, where overall trading activity is 
high (Fig.~\ref{fig:c2_a}), the average wealth does not follow the upward trend 
one would expect in a set-up with ``guaranteed'' average growth of $2\%$ per 
period. In fact, players trade so much that they keep eroding their wealth when 
they sell and affording fewer and fewer number of shares when they buy. This accounts for a negative impact bias on average and 
results in very low earnings at the end of the session, as shown in Fig.~\ref{fig:c2_wt}. On the other hand, in second sessions, where 
overall trading activity is much lower (Fig.~\ref{fig:c3_a}), the average 
wealth does increase with time. 
Nevertheless the number of shares owned eventually 
decreases, which is due not only to excessive trading but also to the fact that 
some subjects cash in their earnings before the end of the experiment and stay out 
of the market from that point onwards. 

This is particularly visible in Fig.~\ref{fig:c3_wd}, when a surge in price triggers selling orders over several 
periods which result in a higher average amount of cash and, naturally, in a 
lower average number of shares. Although this is also visible in the middle of 
the time series in first sessions (Fig.~\ref{fig:c2_wd}), the difference 
between the two cases is that most of the resulting cash is eventually reinvested in the 
first sessions, while in the second sessions 
this does not happen: cash holdings consistently increase after the initial investment phase (Fig.~\ref{fig:c3_wd}).

As we showed in Sec.~\ref{sec:rational_bench} the optimal strategy in our 
experimental set-up would be to collectively buy-and-hold and reap the benefits 
from the baseline average return of $2\%$ per period in the absence of trading. We see that
the behaviour of the agents is very far from this benchmark, even in the second 
sessions, in spite of a significant decrease in activity. The average 
performance in second sessions is indeed still far from what it would have been if 
all subjects used the optimal buy-and-hold strategy, i.e.~despite the 
learning there is still excess trading activity which translates into 
detriment of collective welfare -- since even the ``virtuous'' agents are adversely impacted by the 
trading activity
of excessively ``active'' agents.

\subsubsection{Collective trading modes -- activity correlations, panic \& euphoria}
\label{sssec:collective}

Our subjects trade too much, but can we describe in more detail how correlated their activity is? In fact, our initial intuition
-- that turned out to be quite far from what actually happened --- was that the 
agents would not trade at the beginning of the game, letting the price
rise from its initial value of $100$ to quite high values, say $400$ (EUR 100), before 
starting to worry that others might start selling, pushing the price back down 
and
potentially inducing a panic chain reaction. 
This would have translated into 
either a major crash, or perhaps smaller downward corrections, but in any case
a significant {\it skewness} in the distribution of returns -- absent in principle from the 
bare price series which is constructed to be perfectly symmetrical, since the
noise term in Eq.~\eqref{eq:price_update} is symmetric. 
In fact, as we will see below, the
empirical skewness of the particular realization of the noise turns out to be negative, so 
the reference point that we shall be comparing to must be shifted.

We have therefore measured the relative skewness of the distribution of price changes, 
upon aggregation over time intervals of increasing length, from $\tau=1$ round to $\tau=5$ rounds. 
The idea is that a panic spiral would lead to
a negative skewness that becomes larger and larger when measured on larger time 
intervals, before going back down to zero after the typical correlation time
of the domino effect. This is called the ``leverage effect'' in financial 
markets, and is observed in particular on stock indices where the negative 
skewness
indeed grows as the time scale increases, before decreasing again, albeit very 
slowly \citep{bouchaud2001leverage}.

In order to reduce the measurement noise, it is convenient to measure the 
skewness using two low-moment quantities. One is $1/2-P(r_\tau > m_\tau)$. 
If this quantity is negative, it means that large negative returns
are more probable than large positive returns, as to compensate the excess number of 
positive returns larger than the mean. Another often used quantity is the mean $m_t$ of the
returns minus the median, normalised by the RMS of the returns on the 
same time scale. Again, if the median exceeds the mean, the distribution is 
negatively skewed (see e.g. \cite{reigneron2011principal} for 
further details about these estimators of skewness). Both quantities were found 
to give the same qualitative results,
thus we chose to average these two definitions of skewness and plot them as a 
function of $\tau$, averaged over all first and second sessions.

\begin{figure}[!h]
\centering
	\includegraphics[scale=0.45]{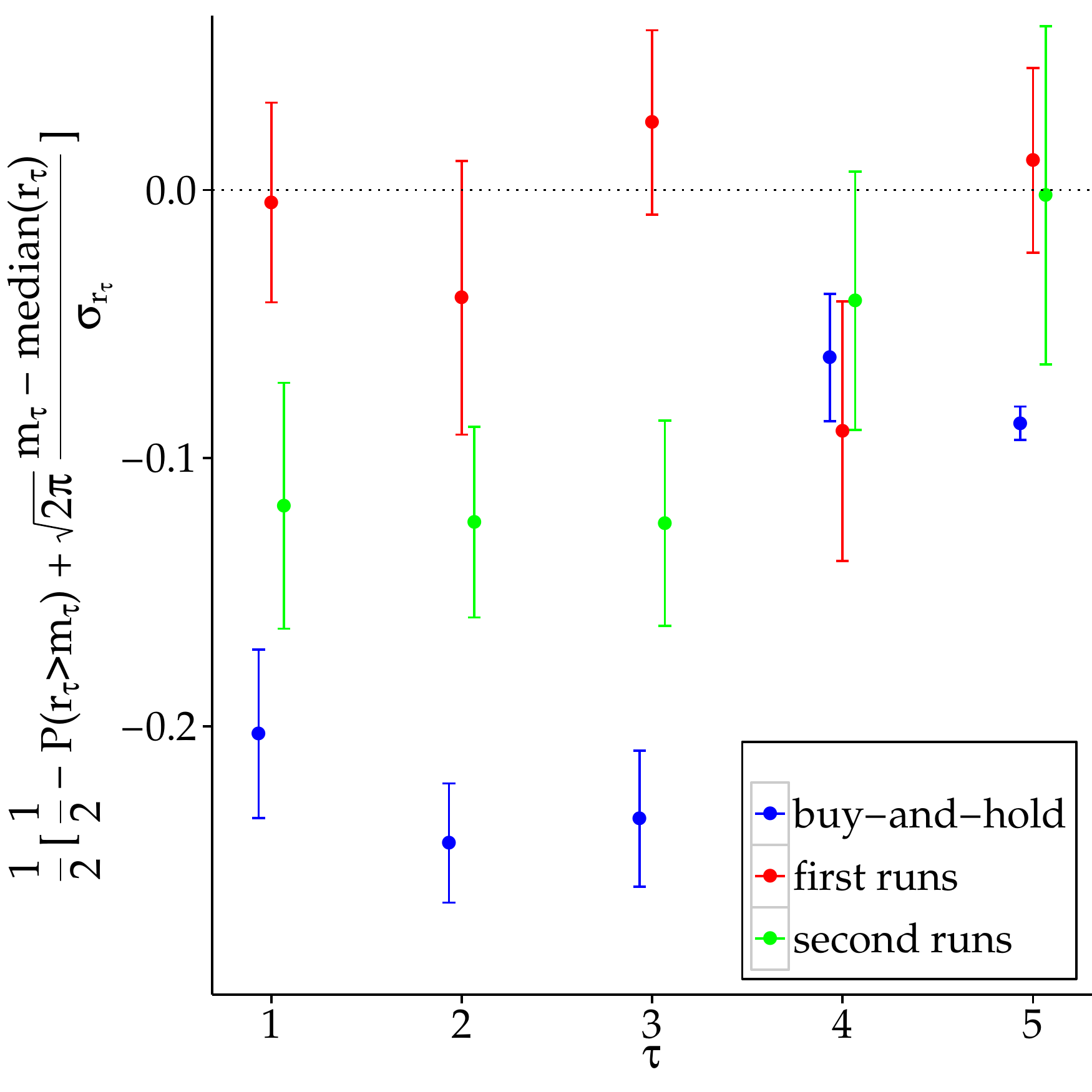}%
	\caption[Content]{Average skewness of price log returns as a function of aggregation length  $\tau$ in first 
sessions (red) and second sessions (green), together with the skewness of log 
returns in the buy-and-hold strategy, i.e. in the absence of trading (blue).}
	\label{fig:skewness}
\end{figure}

The result is shown in Fig.~\ref{fig:skewness}. 
The blue dots correspond exactly to the time series of bare prices 
because there is only one (collective) trade in the buy-and-hold strategy, right at the first period, 
which we discard from the computation. 
Although the bare returns were constructed using a Student's t-distribution with 3 degrees of freedom, 
which by definition is not skewed, we see in Fig.~\ref{fig:skewness} that the bare prices do not have zero skewness. 
This illustrates the role of the noise, which gives way to different values of skewness 
for bare prices depending on the number of periods of the session. We observe in Fig.~\ref{fig:skewness} 
that the realized skewness of trade impacted returns is typically larger (i.e. less negative) than bare returns, but without 
any significant time dependence. This suggests that buying orders tend to be more synchronized than selling orders, specially in the 
first sessions, but that neither buying nor selling orders induce further buy/sell orders. In short, there is no 
destabilising feedback loop in the present setting, which explains why we never 
observed any large crash in our experiments. %; if anything, buy orders tend to be more collective than sell orders.

In order to detect more precisely the synchronisation of our agents, we define an activity correlation matrix ${\bf A}$ as
follows:
\begin{equation}
	\label{eq:coll}
	A_{ij} = \frac{1}{T} \sum_{t} \theta_i(t)\theta_j(t) -  \frac{1}{T} 
\sum_{t} \theta_i(t) \times \frac{1}{T} \sum_{t} \theta_j(t),
\end{equation}
where $\theta_i(t)$ is the activity of agent $i$ at time $t$, $\theta_i(t) = 0$ 
if s/he is inactive, $\theta_i(t) = \pm 1$ if s/he buys or sells.

\begin{figure}[!htpb]
\centering
	\includegraphics[scale=0.3]{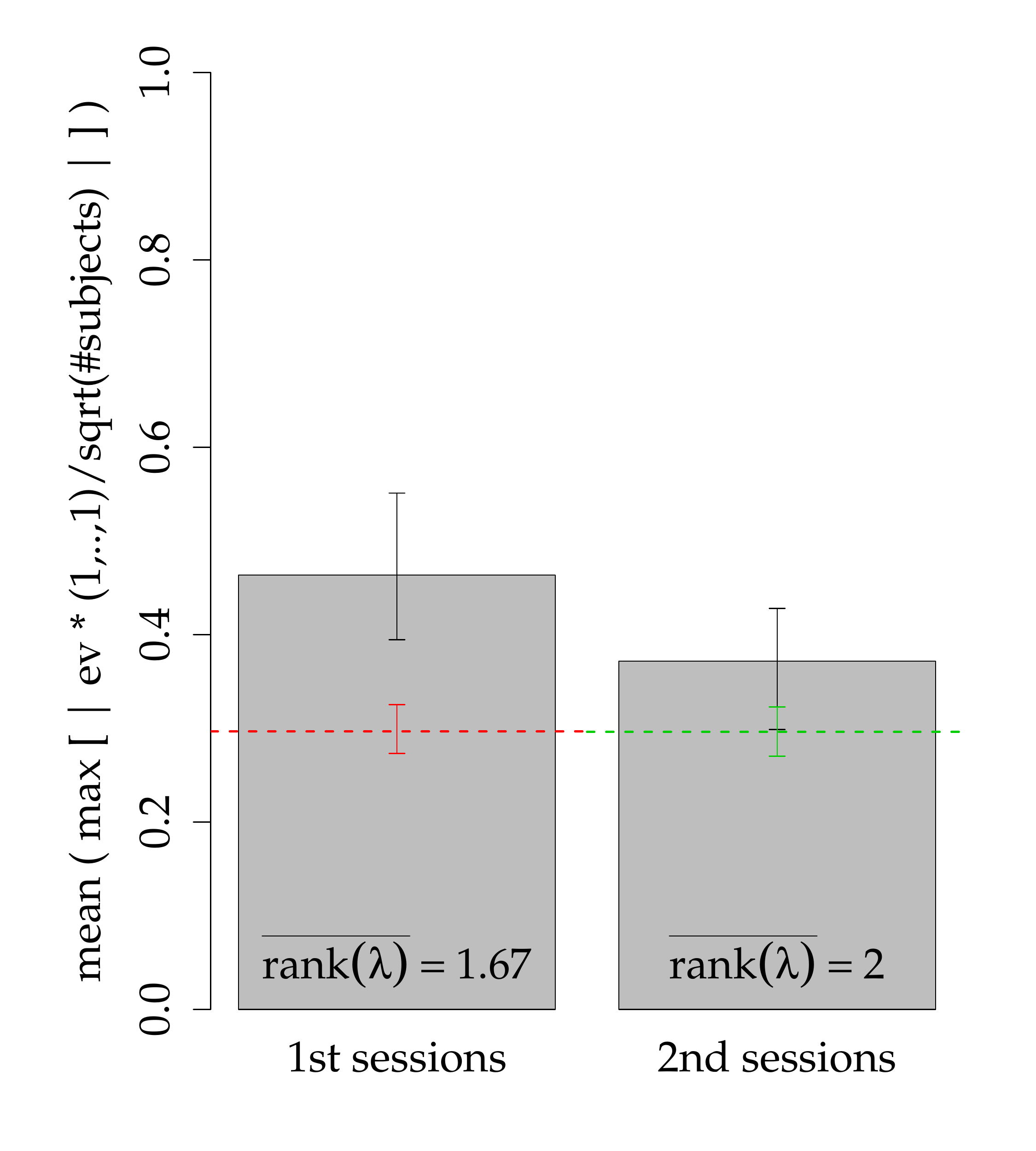}%
	\caption[Content]{Average maximum absolute value of the three dot products between 
	the eigenvectors corresponding to the three largest eigenvalues, and the unit vector, with the corresponding statistical error bars. 
All orders are considered except for the first time step, in which we expect a 
natural bias towards synchronization. The dashed horizontal line around $0.3$ corresponds to the null hypothesis of completely
uncorrelated actions.  }
	\label{fig:c2c3_coll}
\end{figure}

For each session, we diagonalize ${\bf A}$ and study the three largest 
eigenvalues, corresponding to the more important principal components of the 
subjects'
activity. In order to detect synchronisation, where a substantial fraction of 
agents tend to act in exactly the same way across the experiment, we compute 
the absolute value of the dot
products of these three eigenvectors $\vec v_1, \vec v_2, \vec v_3$ and the 
uniform vector $\vec e= (1,1, \dots, 1)/\sqrt{N}$. Then, we average the maximum of these three numbers over all runs. 
It may indeed happen that the ``synchronized'' mode does not correspond to the largest
eigenvalue of ${\bf A}$, while still being amongst the three most important ones, and the above procedure
allows one to capture these cases. The 
resulting values are represented in Fig. ~\ref{fig:c2c3_coll} for the first and 
second sessions, and compared with a null-hypothesis benchmark, obtained using 1000 random bootstrap replicates of the experiments. 

The dashed lines depict the cases where agents would act completely at random, which would lead to a value of this average 
maximum overlap at approximately $0.35$. 
We see that the experimental
results are clearly larger than the benchmark case: approximately $0.57$ for the 
first sessions and approximately $0.5$ for the second sessions (compared to a maximum
value of $1$ for a fully collective activity mode).

The method above allows us to make a quantitative analysis and statement about 
overall synchronization in our experiment, be it through selling or buying 
orders, and we find that there is indeed significant synchronisation. 

We also consider the separate cases of synchronization for selling 
orders and for buying orders. In order to do this, we construct an activity 
correlation matrix as in Eq.~\eqref{eq:coll} but change the definition of 
$\theta_i(t)$ accordingly. 
This way, when we study the synchronization 
concerning only buying orders, we define 
$\theta_i(t) = 0$ if agent $i$ is inactive or sells and $\theta_i(t) = 1$ if he 
buys. Likewise, in the case where we look into synchronization over selling 
orders only, we set $\theta_i(t) = 0$ if 
agent $i$ is inactive or buys and $\theta_i(t) = -1$ if he sells. 

We see in 
Fig.~\ref{fig:c2c3_coll_buy} and in Fig.~\ref{fig:c2c3_coll_sell} in Appendix \ref{app:extra_figs} that 
splitting the data set as explained does yield similar results: the experimental
results are larger than the benchmark for each case, with the difference more 
marked in first sessions than in second sessions. Again, the synchronisation of 
buy orders appears to be, according to this metric, slightly stronger than that of sell orders.

We therefore conclude that although our subjects cannot directly communicate 
with one another, there is a significant synchronisation of their activity, in 
particular during the first sessions and, as the skewness of the distribution 
reveals, for the buying activity. The mechanism for this synchronisation can
only come from the common source of information that the subjects all observe, 
namely the price time series itself -- see below for more about this.

Furthermore, we observe an asymmetry if we repeat the above method conditional on the sign of previous returns, in the sense that synchronisation is stronger for buying orders conditional on negative previous returns and for selling orders conditional on positive previous returns. This indicates some sort of ``mean reversion'', which is in line with the findings on subjects' trading behaviour discussed in Subsection \ref{ssec:predictions} below.

\subsubsection{Clustering}\label{sssec:clusters}

Fig.~\ref{fig:activity} shows us, once again, that the average final wealth in first sessions 
is much smaller than in second sessions, which is tied to the higher average 
trading activity of the subjects when they play the game for the first time. In 
second sessions, we observe a number of subjects who kept trading activity very low, increasing their chances of a positive payout at the end of the 
experiment. As we discussed in Sec.~\ref{sssec:wealth}, this is an indication 
that the subjects learn. In any case, there are always traders who keep trading 
at very high rates and lose money in the process.

However, Fig.~\ref{fig:activity} does not provide insight about common patterns 
in the behaviour of the subjects. We know from Fig.~\ref{fig:positions} that at 
least in second market sessions a number of subjects use the buy-and-hold strategy or 
similar, which corresponds to the green horizontal ``corridors'' in the figure. 
Therefore, we apply clustering techniques to search for groups of subjects with similar trading profiles in  
the data sets. Afterwards, we look into the trading activity and trading performance in each cluster.

As in \cite{tumminello2011statistically} and \cite{tumminello2012identification}, we applied false discovery rate (FDR) methods to validate 
links between subjects to the data set with composite data from first sessions 
and to the data set with composite data from second sessions. The variable used 
to establish links (i.e. similarity) between subjects was their position -- in or out of the market -- over time for 
each subject. The FDR rejection threshold was $1\%$. 

\begin{figure}[!htpb]
\centering
	\subfloat[First sessions\label{fig:c2_clusters}]{%
		\includegraphics[scale=0.6]{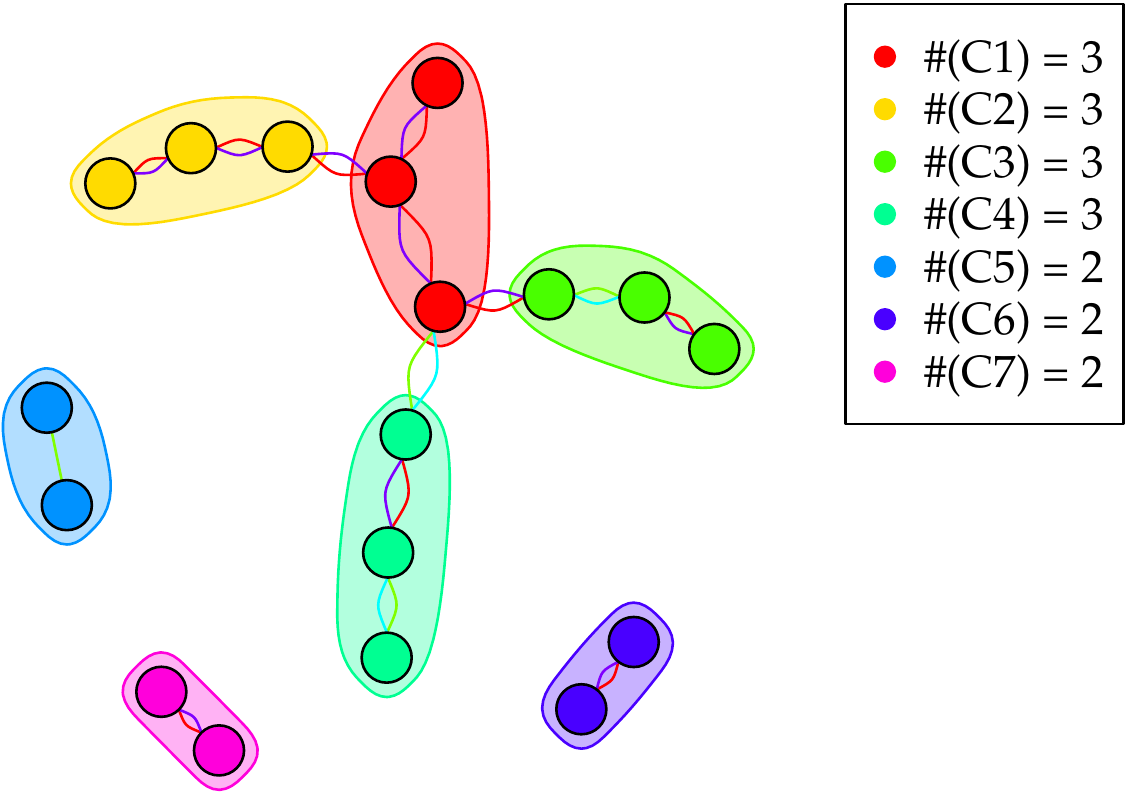}%
		}\hfill
	\subfloat[Second sessions\label{fig:c3_clusters}]{%
		\includegraphics[scale=0.6]{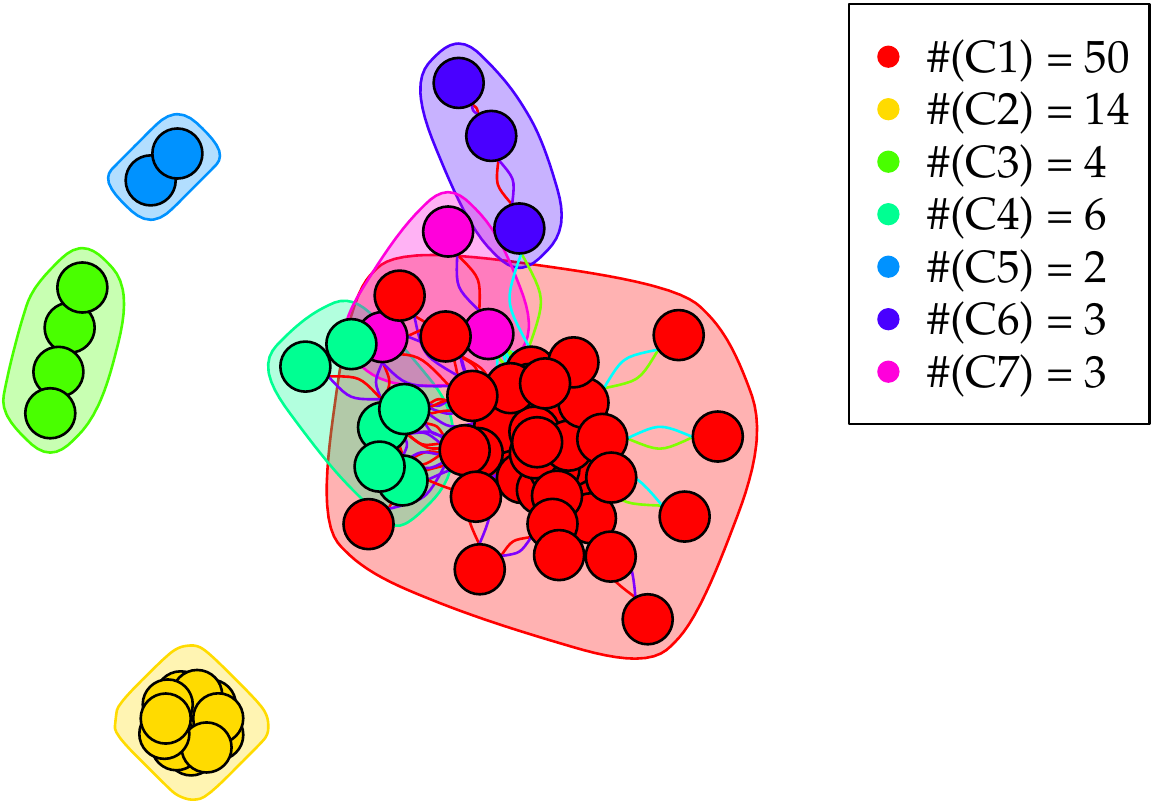}%
		}\hfill
	\caption[Content]{Clusters for first and second sessions using the FDR 
algorithm with a threshold of $1\%$ applied to positions -- in or out of the 
market -- over time. The number of subjects in each cluster is displayed in the legends.}
	\label{fig:clusters}
\end{figure}

The clusters are visualised in Fig.~\ref{fig:clusters}, which also displays the number of subjects in each cluster for first sessions and second sessions. 

	\begin{figure}[!h]
	
	\subfloat[First sessions\label{fig:clusters_c2_actfw}]{%
		\includegraphics[scale=0.45]{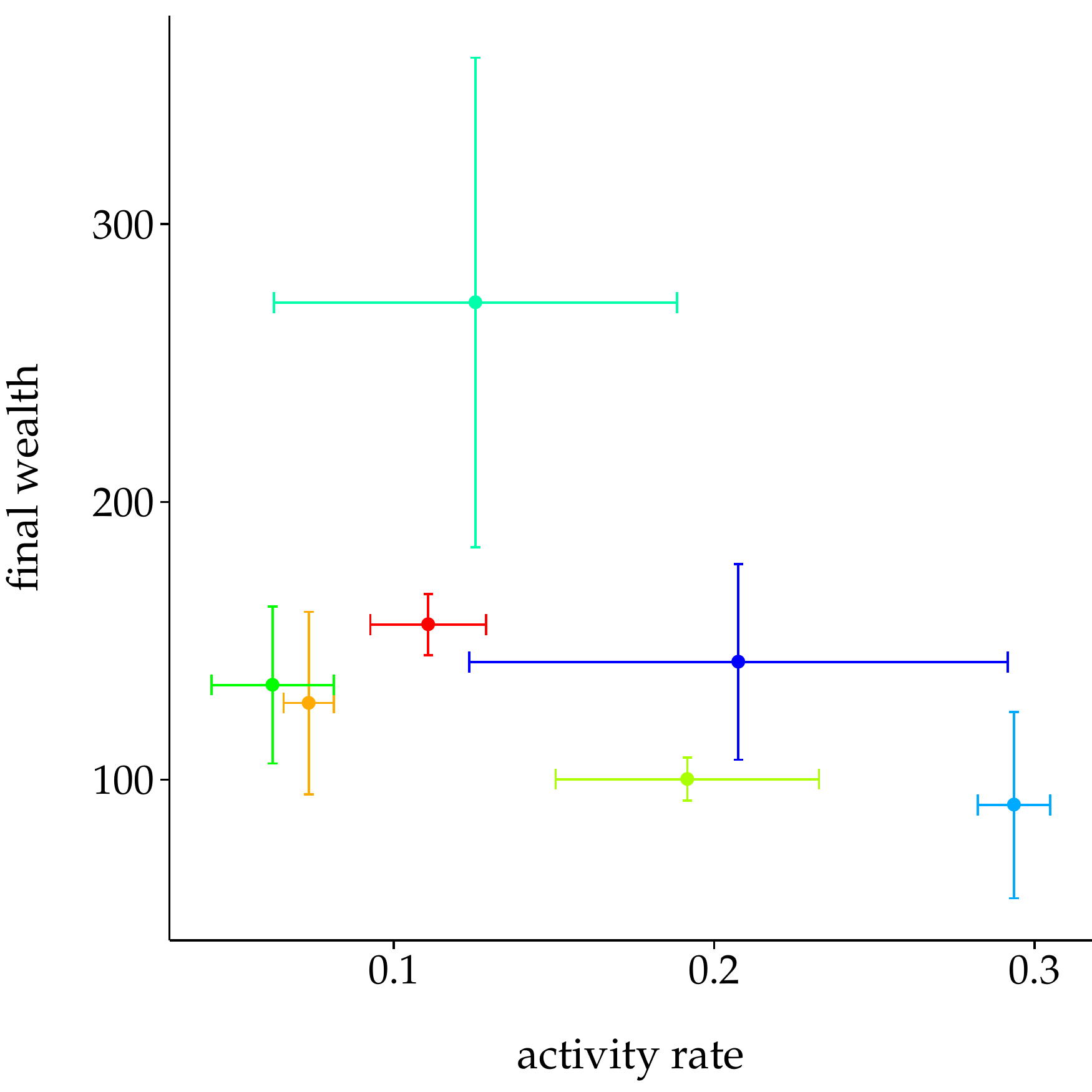}%
		}\hfill
	\subfloat[Second sessions\label{fig:clusters_c3_actfw}]{%
		\includegraphics[scale=0.45]{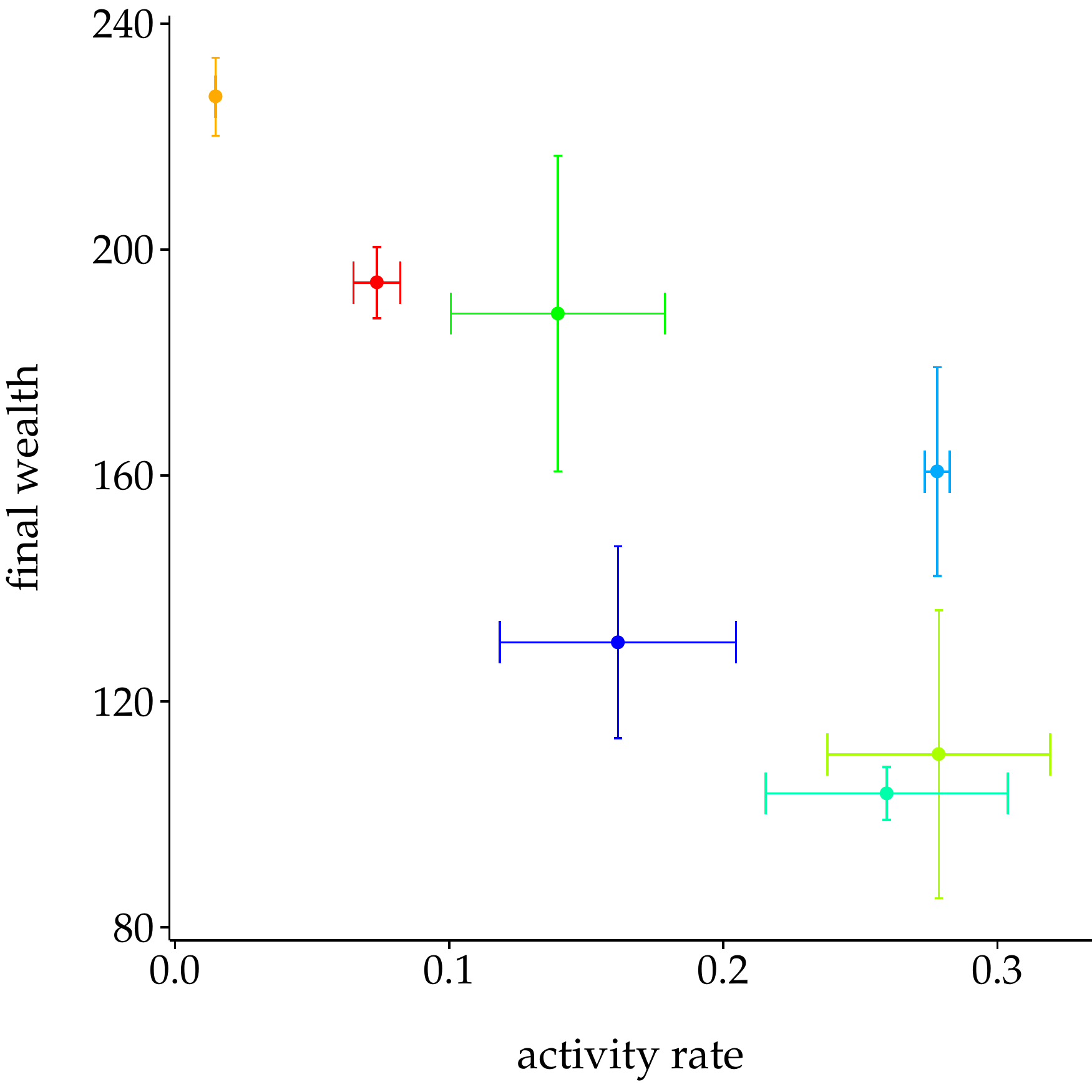}%
		}\hfill
	\caption[Content]{Average trading activity and final wealth for each cluster represented in Fig. \ref{fig:clusters}.}
	\label{fig:clusters_actfw}
\end{figure}
The clusters 
identified in the second sessions are much larger than those identified in first 
sessions, which is expected because the number of ``intermittent'' players in 
the game was lower in second sessions.

We see in Fig.~\ref{fig:clusters_actfw} that
clusters with a lower average trading activity tend to have a higher average final wealth. A notable example
is the cluster number $2$ in Fig.~\ref{fig:clusters_c3_actfw}, which includes $14$ subjects (see Fig.~\ref{fig:c3_clusters}) who kept trading to a minimum in second runs and maximized their returns. Conversely, the cluster number $6$ in Fig.~\ref{fig:clusters_c2_actfw} consists of $2$ traders (see Fig.~\ref{fig:c2_clusters}) with very high average trading activity and, as a consequence, low final wealth. 

\subsection{Risk attitude and activity rate}
\label{ssec:risk}
To measure risk aversion of subjects we use the paired lottery choice instrument of \cite{holt2002risk}.
The Holt-Laury paired-lottery choice
task is a commonly-used individual decision-making experiment for measuring individual risk
attitudes. This second experimental task was not announced in advance; subjects were instructed
that, if they were willing, they could participate in a second experiment that would
last an additional 10-15 minutes for which they could earn an additional monetary payment. All subjects agreed to participate in this second
experiment.
In this task subjects choose between a lottery with high variance of
payoffs (Option B) and lottery with less variance (Option A). 
As in \cite{holt2002risk} we use the relative frequency of B-choices (``risky'' choices) as a measure for a preference for risk. Moreover, in order to make the amounts at stake comparable to what subjects
could earn over an average session of trading periods and to assess whether or not the risk attitudes of subjects depended on the wealth levels involved, we elicited subjects' choices for two lotteries, corresponding to two times ($2\times$) and ten times ($10\times$) the amounts offered by \cite{holt2002risk} in their baseline treatment. Subjects were ex-ante informed that they would throw a dice to determine which of the two lotteries would determine their payoff from this additional experimental task. 

Appendix \ref{sec:instructions} includes the instructions for the Holt-Laury paired-lottery choice
experiment.

\begin{figure}[!h]
\centering
	\includegraphics[scale=0.45]{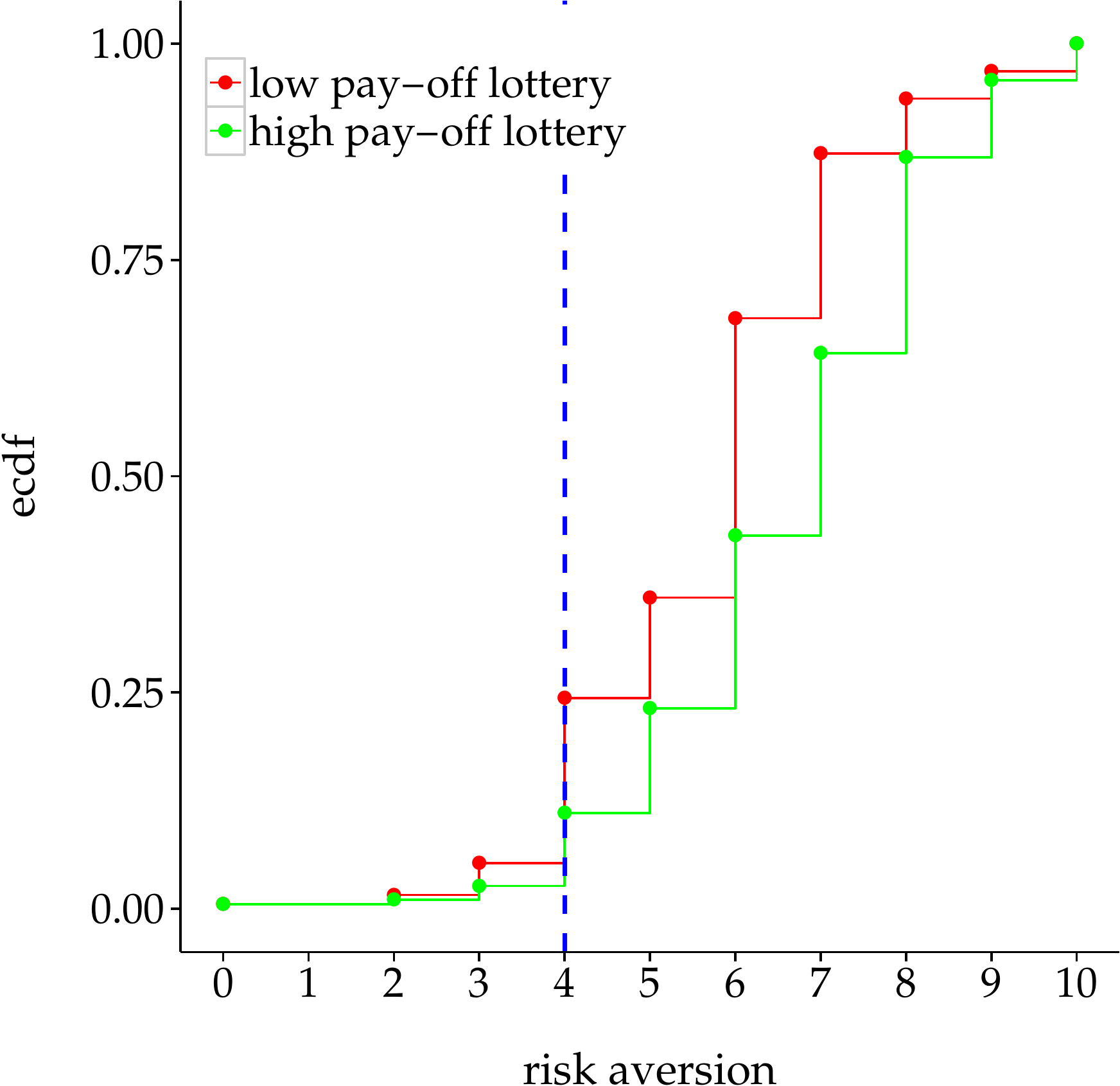}%
	\caption[Content]{Risk aversion cumulative distribution function. The x-axis shows the number of A-choices (``safe choices'') in the paired lotteries. The higher this level, the more risk-averse is the agent. The blue vertical line shows where a risk neutral subject would be.}
	\label{fig:risk_ecdf}
\end{figure}

In Fig.~\ref{fig:risk_ecdf} we show the distribution of subjects
according to their risk aversion, both for the $2\times$ and the $10\times$ lottery.\footnote{We discarded from our data set the cases in which the subjects chose the safe lottery after having
previously chosen the risky option for a lower pay-off advantage.}

The blue vertical line shows where a risk neutral subject
would be, based on the expected pay-off differences alone. The fact that the majority
of the subjects -- $76\%$ for the low-pay-off and $89\%$ for the high pay-off lotteries -- 
have a number of ``safe'' A-choices larger than $4$ indicates that, overall, the participants in our experiment 
were risk-averse. Moreover, if we compare the two curves, red and green, we 
see that the subjects tend to safer choices when the lottery pay-off is higher, 
which indicates that the risk aversion of our population not only depends but 
increases with the pay-off level. This is corroborated by the Welch two-sample t-test,
which states that the mean risk aversion values for low pay-off and high pay-off
are statistically different with a p-value of $1.5 \, 10^{-6}$. We also estimate the parameters of the power-expo utility function in Eq.~\eqref{eq:utility} using the lottery choices of our pool of subjects. As in \cite{holt2002risk}, we find evidence for increasing relative risk aversion
and decreasing absolute risk aversion, i.e. positive estimates for both parameters $\alpha$ and $r$. Details on the estimation procedure and results are reported in Appendix \ref{app:risk_est}.

We now relate individual risk attitude, as elicited by the binary lottery choices, to trading
behaviour. Previous experimental studies on the relation between elicited risk attitudes and aggregate market behaviour \citep{robin2012bubbles,fellner2007risk} show that the higher the degree of risk aversion the lower the observed market activity. On the other hand, \cite{michailova2010overconfidencerisk} finds no significant effect of the number of safe choices in paired binary lotteries on the frequency of trading.

\begin{figure}[!htpb]
\subfloat[\label{fig:risk_tr_w}]{%
\includegraphics[scale=0.45]{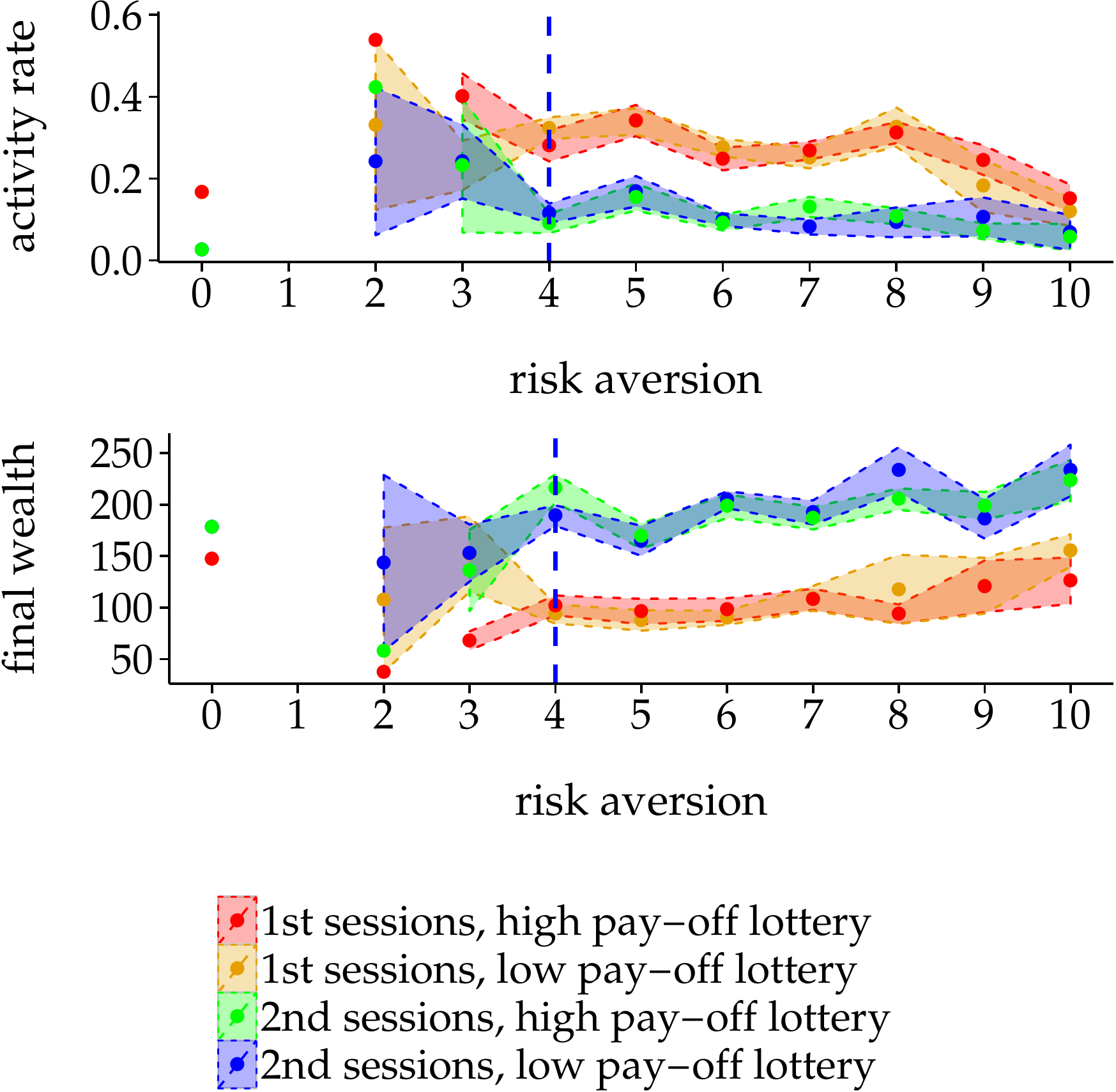}%
}\hfill
\subfloat[\label{fig:risk_tr_w_lm}]{%
\includegraphics[scale=0.45]{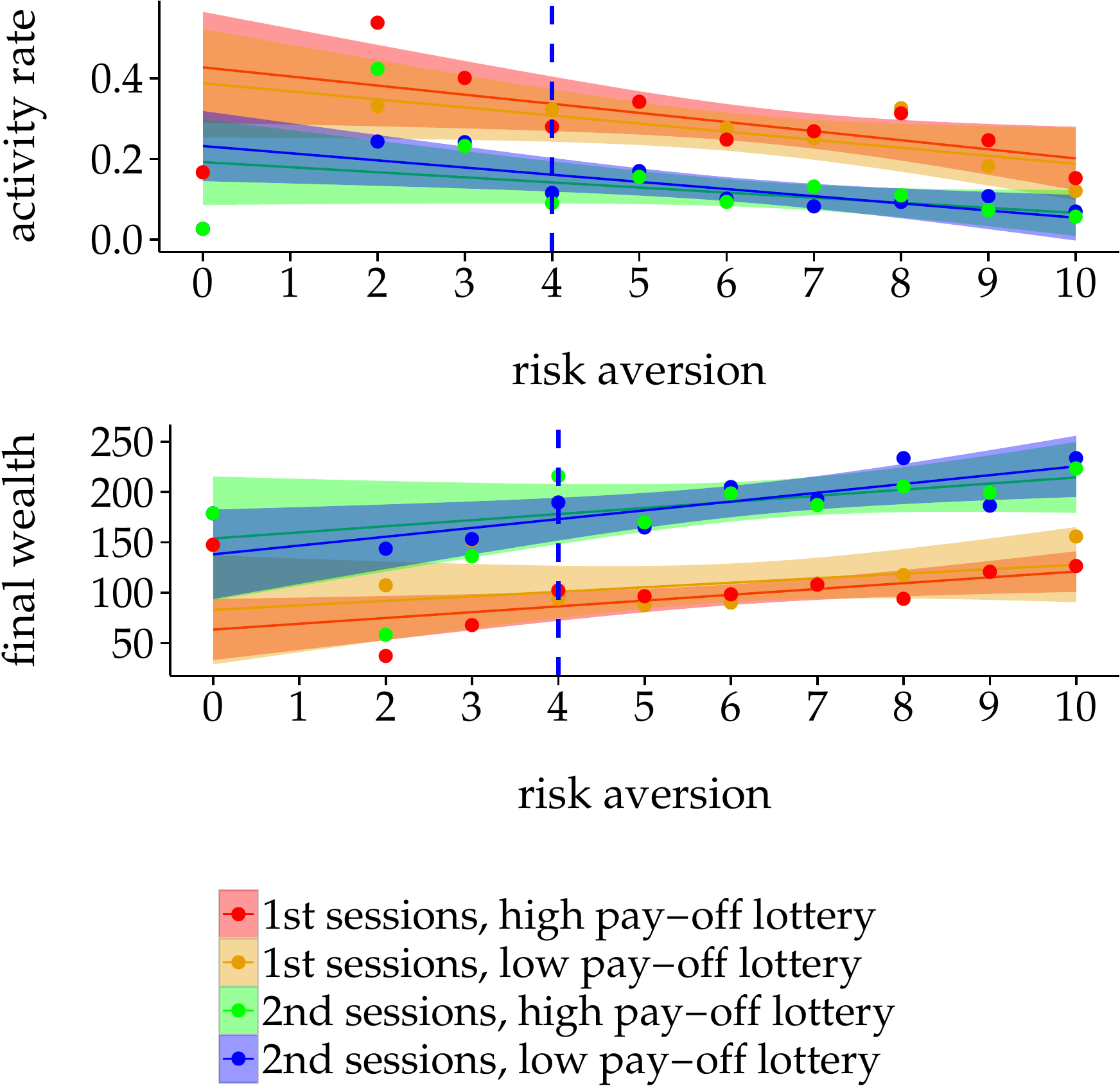}%
}\hfill
\caption[Content]{Risk attitudes, trading activity and final wealth. Plot (a) displays average activity rate and final wealth for each level of risk aversion. Plot (b) shows linear regressions with $95\%$ intervals of confidence of the average activity rate and final wealth on risk aversion. Clearly, more risk averse subjects trade less and end up with a higher final wealth.}
	\label{fig:risk_act-fw}
\end{figure}

Fig.~\ref{fig:risk_act-fw} displays average activity rates and levels of final wealth for different risk attitudes in both first and second runs. 

Our results clearly show that the activity rate decreases and the final wealth increases with the level of risk aversion, both in first runs (red and orange) and in second runs (green and blue). In other words, preference for risk is an important determinant of excess trading in our experiment. Our findings are in line with the empirical evidence suggesting that risk-loving, overconfident individuals are more willing to invest in stocks \citep{Keller2006Investing} and engage in speculative activity \citep{odeanboys,camacho2012investment}.

\subsection{Price forecasts and trading behaviour}
\label{ssec:predictions}

The fact that the subjects input their price predictions throughout the experiment allows us to have a glimpse of their frame of mind. In fact, trades only tell about the consequence of the state of mind (i.e. the price expectation) of traders when they are active. But traders (both in real life and in experiments) are in fact inactive most of the time. As a consequence, trades alone are unlikely to be able to explain why traders are inactive. Since we have both trades and subjects' price expectations, we are able to give a consistent picture of activity and inactivity as a consequence of price return expectations.

In both market sessions, the subjects did not input anything in about $7\%$ of the time, as price prediction was not a mandatory activity (although monetarily incentivised); in the following, we restrict our analysis to the subjects that did report their predictions.

\begin{figure}[!htpb]
\centering
	\includegraphics[scale=0.43]{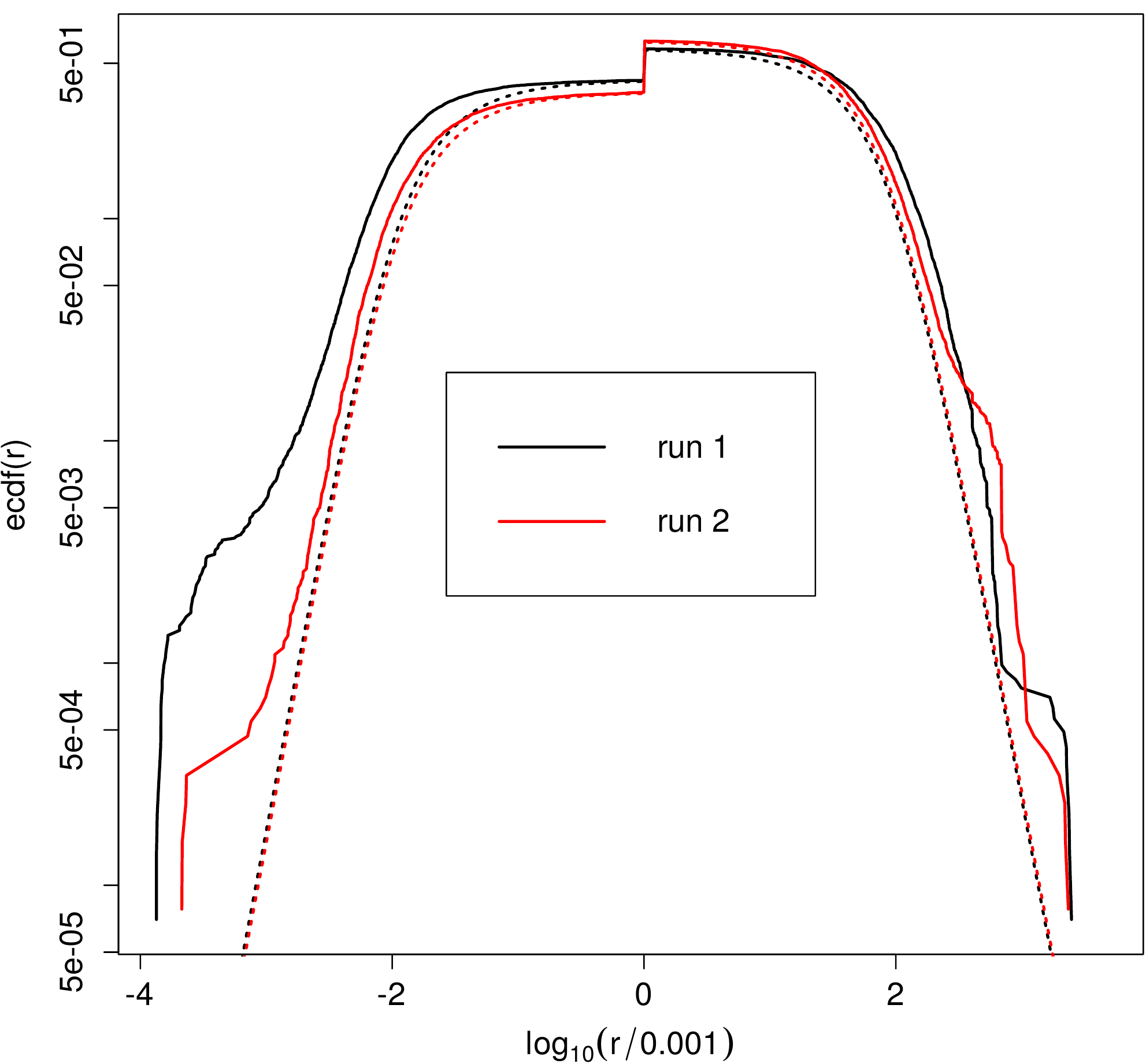}	
	\caption[Content]{Reciprocal empirical cumulative distribution functions of negative (left) and positive (right) expected price returns during the first and second sessions (black and red lines, respectively). The baseline return distribution is plotted in dashed lines. The jump at $r=0$ indicates the respective fraction of positive and negative predictions.}
	\label{fig:prediction_posneg}
\end{figure}

The discussion focuses on the predicted log returns, i.e.~from subject $i$'s price predictions $\widehat{p_i}(t+1)$, we compute the predicted log returns $\widehat{r_i} (t+1)=\log [\widehat{p_i}(t+1)/p(t)]$, for all subjects. The average prediction in the first market sessions is $-0.01$, and $+0.02$ in the second market sessions, i.e.~exactly in line with the average return $m=2\%$ in absence of trading. The percentage of positive predicted returns is 54\% in the first run, and 58\% in the second run. Fig.~\ref{fig:prediction_posneg} illustrates the full empirical cumulative distribution functions of expectations for both runs and for positive and negative return separately. 

The starting point for each of the positive and negative distributions represents the fraction of positive and negative expectations, thus the higher jump at  $r=0$ for the second session reflects the increase in the fraction of positive expected returns. 

We then check how the distributions of positive and negative expected returns are related to the baseline return distribution, determined by the Student noise term $s\eta_t$ in the price updating rule. The baseline return distribution is plotted in dashed lines in Fig.~\ref{fig:prediction_posneg}; comparison between the latter and empirical distribution of expected returns gives a first clue about the type of extrapolation rules from past returns that the agents use. The asymptotic power-law tails of Student's t-distributed variables, such as the noise $\eta_t$,  are unchanged under summation, by virtue of the central limit theorem. This implies that linear expectations yield expectation distributions with power-laws tails that have the same exponent. On the other hand, panic or euphoria may lead to non-linear extrapolations and thus may modify either the tail exponent of these distributions, or even the nature of the tails.

The most obvious finding is that the actions of the agents increase the volatility of the baseline signal (in dashed lines) as the empirical distribution functions are above the baseline signal for both sequences. The amplification of the noise for positive expectations is almost the same in the two sequences, while for negative expectations there are marked differences between the two runs as the scale of negative expectations was much larger during the first run. 
We used robust power-law tail fitting \citep{clauset2009power,poweRlaw} and  determined the most likely starting point of a power-law $r_{min}$ and the exponent $\alpha$ (see Table \ref{table:rminalpha}).\footnote{The parameters $r_{min}$ and $\alpha$ for the power-law are not to be confused with the parameters of the power-expo utility function in Eq.~\eqref{eq:utility}.} Quite remarkably, the parameters of the positive and negative tails are simply swapped between the two runs: thus not only the scale of negative expectations changes, but the nature of largest positive and negative expectations also changes. The fitted tail exponent is not far from $3$, the one of the Student noise showing once again the absence of destabilizing feedback loops.
 
\begin{table}[!htpb]
\centering
\begin{tabular}{|c|c|c|}
\hline 
  & $r_ {min}$ & $\alpha$ \\ 
\hline 
run 1 $r<0$ & 0.21 & 3.5 \\ 
\hline 
run 2 $r>0$ & 0.10 & 2.7 \\ 
\hline 
\hline 
run 1 $r<0$ & 0.13 & 2.6 \\ 
\hline 
run 2 $r>0$ & 0.18 & 3.5 \\ 
\hline 
\end{tabular} 
\caption{Fits of the power-law part of return expectations; $r_{min}$ denotes the most likely starting point of the power-law.\label{table:rminalpha}}
\end{table}

The fact that the subjects have heavy-tailed predictions suggests that they form their predictions by learning from past returns, which do contain heavy tails because of the Student's t-distributed noise. We thus hypothesize some relationship between predicted returns and past returns. This is in line with the best established fact about real investors, which is the contrarian nature of their trades: their net investment over a given period is anti-correlated with past price returns \citep{jackson2003aggregate,kaniel2008individual,grinblatt2000investment,challet2013robust}. In addition, previous experiments \citep{hommes2005coordination} have
demonstrated that four simple classes of linear predictors using past returns are usually enough to reproduce the observed price dynamics. 

Based on the considerations outlined above, we first the return predictions of each subject with a linear model.
\begin{equation}\label{eq:w0w1}
\widehat{r} (t+1)=\omega_{0}+\omega_{1} r(t),
\end{equation}
where $\omega_{0}$ and $\omega_{1}\in \mathbb{R}$. Price return predictions are fitted separately for each trader, each session, and each possible action. We fit $\omega_0$ and $\omega_1$ simultaneously. Sec.~\ref{sssec:w_0} discusses results for $\omega_0$ while Sec.~\ref{sssec:w_1} is devoted to $\omega_1$.

\subsubsection{Average predictions ($\omega_0$)}
\label{sssec:w_0}
\begin{figure}[!htpb]
	\includegraphics[scale=0.45]{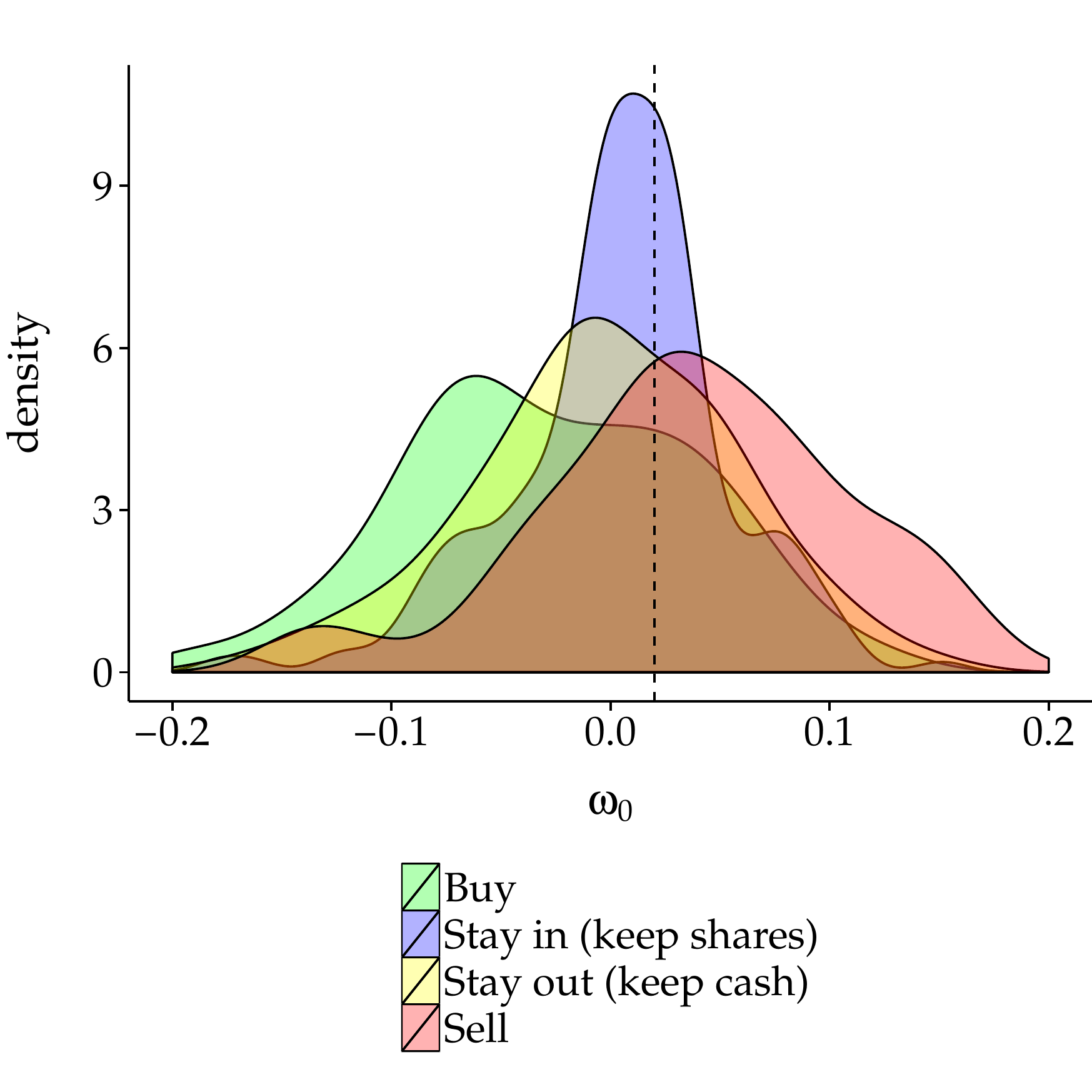}
	\includegraphics[scale=0.45]{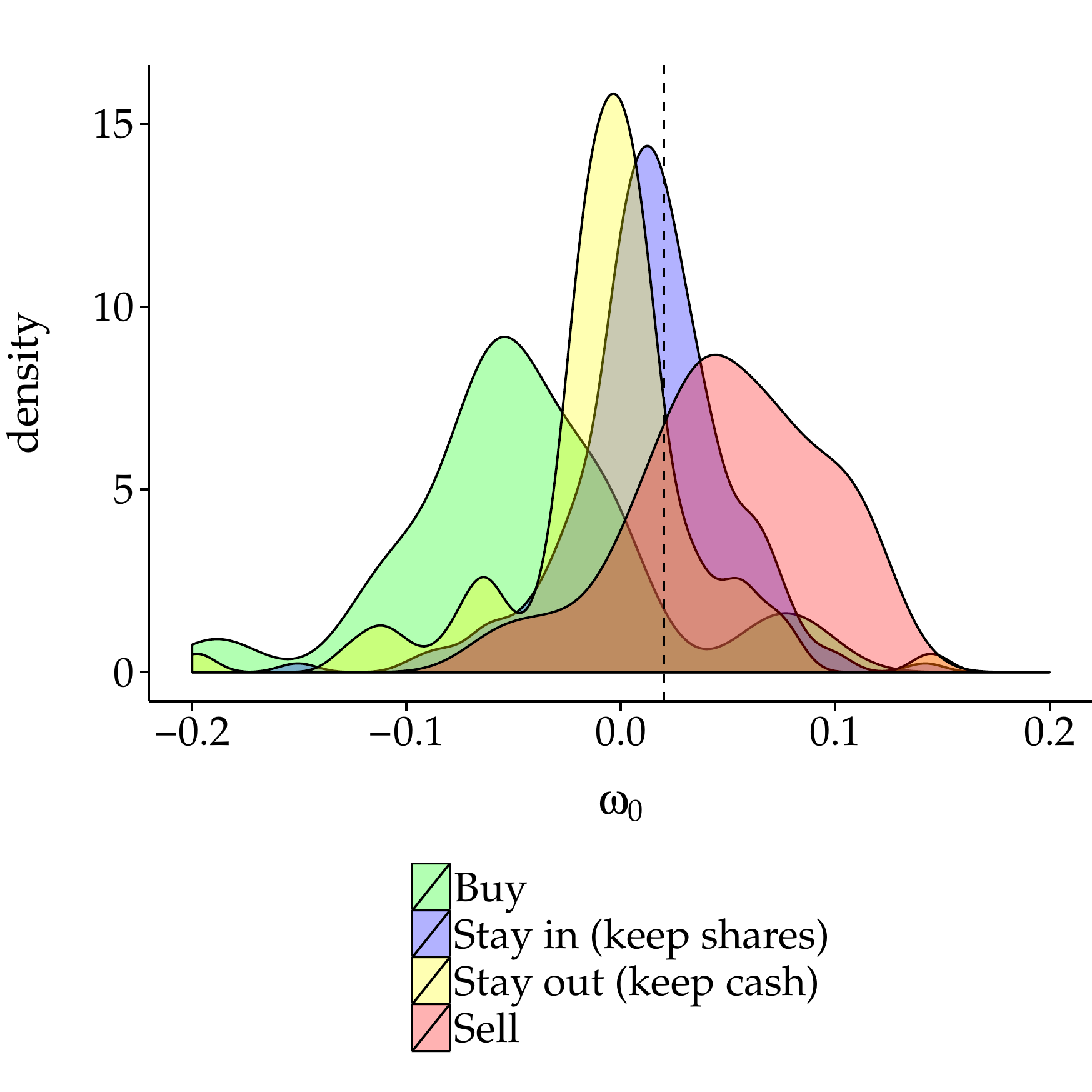}
	\caption[Content]{Densities of trader average return prediction $\omega_0$ during the first sessions (left plot) and second sessions (right plot) for the four types of decisions. Dashed vertical lines refer to the baseline return of $2\%$.}
	\label{fig:w_0}
\end{figure}

In order to give a picture of traders' movements on the market we compute return expectations conditionally on the actions of the subjects. There are four possible actions: buying, selling, holding shares and holding cash. Fig.~\ref{fig:w_0} reports the conditional distributions of price return predictions, for both sessions. 

The results are qualitatively the same for both market sequences: the conditional distributions are clearly separated, as shown in Tab.~\ref{table:ttests_w0} below; the main difference between the two runs is that the variance of expectations among the population is much reduced in the second run. 

\begin{table}[!htpb]
\centering
\begin{tabular}{|c|c|c|c|}
\hline 
$\omega_0$, 1nd run  & Sell & Hold cash & Hold shares \\ 
\hline 
Buy &   $1.5\,10^{-8}$ & $3.9\,10^{-3}$ & $1.2\,10^{-5}$ \\ 
\hline 
Sell &    & $3.8\,10^{-5}$ & $8.2\,10^{-5}$ \\ 
\hline 
Hold cash   &  &  & $1.5\,10^{-1}$ \\ 
\hline 
%Hold shares &  &  &  &  \\ 
%\hline 
\end{tabular} \\ \vspace{0.2cm}
\begin{tabular}{|c|c|c|c|}
\hline 
$\omega_0$, 2nd run & Sell & Hold cash & Hold shares \\ 
\hline 
Buy &   $2.1\,10^{-7}$ & $1.5\,10^{-4}$ & $1.0\,10^{-7}$ \\ 
\hline 
Sell &    & $6.1\,10^{-7}$ & $5.9\,10^{-5}$ \\ 
\hline 
Hold cash   &  &  & $6.6\,10^{-6}$ \\ 
\hline 
%Hold shares &  &  &  &  \\ 
%\hline 
\end{tabular}
\caption{Tests of the difference of distributions of $\omega_0$ among the subjects, conditional on two given actions. The table reports the p-values of Mann-Whitney tests for each possible pair of actions.}
\label{table:ttests_w0}
\end{table}

Let us break down the results for each possible action:
\begin{enumerate}
\item When the subjects hold assets, their expectations are in line with the baseline return of 2\%.
\item When the subjects hold cash, their expectations are significantly lower (essentially zero).
\item When the subjects make a transaction, however, their expectations of the next returns are anti-correlated with their actions, i.e., they buy when they expect a negative price return and vice versa.
\end{enumerate}

Thus, the actions of the subjects are fully consistent with their expectations: they do not invest when they do not perceive it as worthwhile and they keep their shares when they have a positive expectation of future gains. The actions of trading subjects are instead consistent with a \textit{``buy low, sell high''} strategy. In fact, knowing that transactions will be executed at the price realized in the next period, subjects submit buy orders when they expect negative returns and sell orders when they expect positive returns.

\subsubsection{Predictions and past price returns ($\omega_1$)}
\label{sssec:w_1}
\begin{figure}[!htpb]
	\includegraphics[scale=0.45]{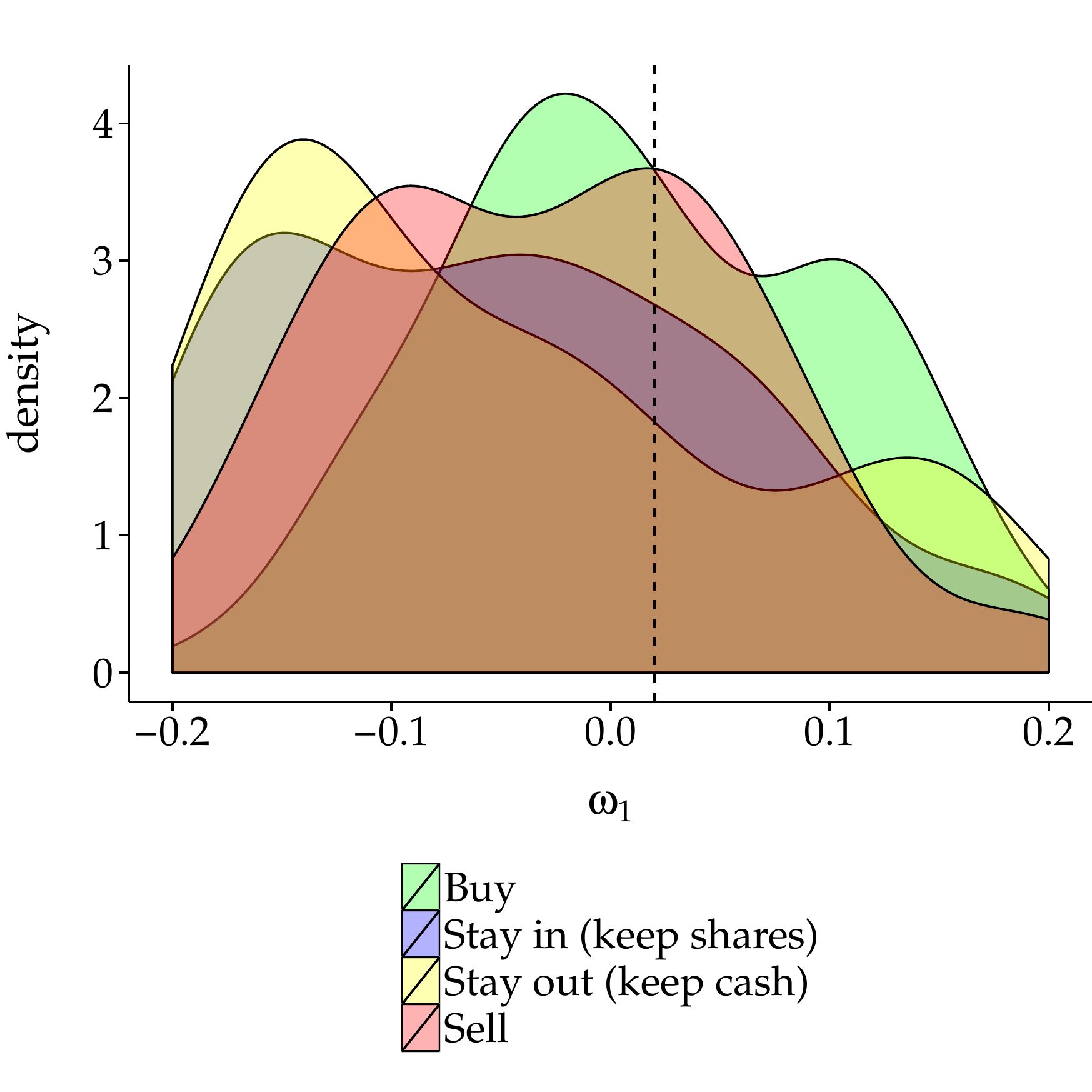}
	\includegraphics[scale=0.45]{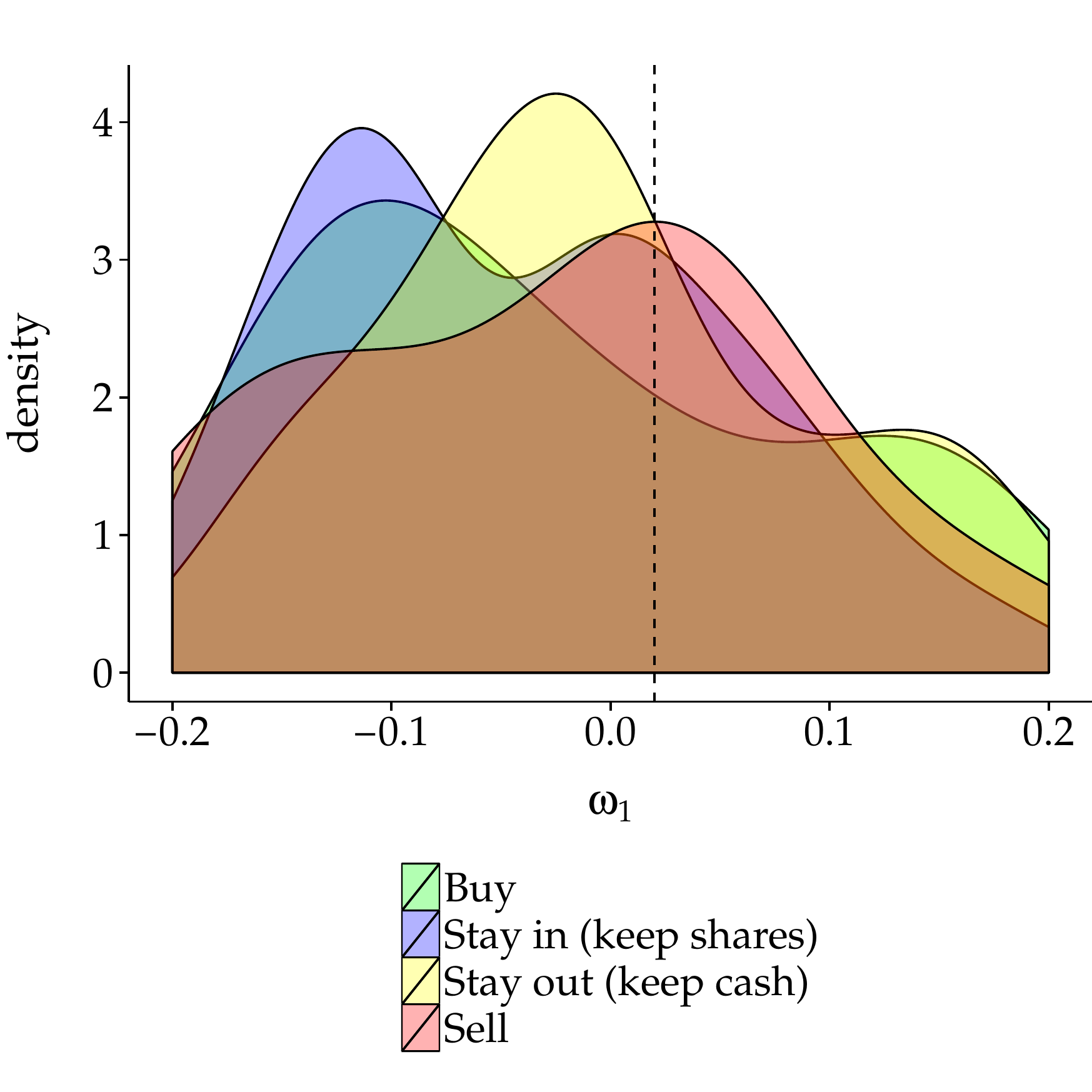}
	\caption[Content]{Densities of trader conditional return prediction $\omega_1$ during the first sessions (left plot) and second sessions (right plot) for the four types of decisions. Dashed vertical lines refer to the baseline return of $2\%$.}
	\label{fig:w_1}
\end{figure}

We find that the importance of the coefficient $\omega_1$, which encodes the linear extrapolation of the past return on future returns, is very weak. Figure \ref{fig:w_1} reports the conditional distributions of past returns' impact coefficient for both runs, while Tab.~\ref{table:ttests_w1} reports the p-values of the Mann-Whitney tests between coefficients $\omega_1$ between all state pairs for all subjects. 
\begin{table}[!htpb]
\centering
 \begin{tabular}{|c|c|c|c|}
\hline 
$\omega_1$, 1st run & Sell & Hold cash & Hold shares \\ 
\hline 
Buy &   $9.6\,10^{-2}$ & $2.3\,10^{-5}$ & $8.8\,10^{-10}$ \\ 
\hline 
Sell &    & $3.7\,10^{-2}$ & $1.7\, 10^{-4}$\\ 
\hline 
Hold cash   &  &  & $2.9\,10^{-2}$ \\ 
\hline 
%Hold shares &  &  &  &  \\ 
%\hline 
\end{tabular}\\ \vspace{0.2cm}
\begin{tabular}{|c|c|c|c|}
\hline 
$\omega_1$, 2nd run  & Sell & Hold cash & Hold shares \\ 
\hline 
Buy &  $6.2\,10^{-1}$ & $7.6\,10^{-1}$ &$4.1\,10^{-2}$ \\ 
\hline 
Sell &    & $5.6\,10^{-1} $& $8.6\,10^{-2}$ \\ 
\hline 
Hold cash   &  &  & $2.9\,10^{-4} $\\ 
\hline 
%Hold shares &  &  &  &  \\ 
%\hline 
\end{tabular}
\caption{Tests of the difference of distributions of $\omega_1$ among the subjects, conditional on two given actions. The table reports the p-values of Mann-Whitney tests for each possible pair of actions.}
\label{table:ttests_w1}
\end{table}

The plot shows that during the first run, this coefficient was negative when the agents did not act and zero when they did trade. The second run is different: the coefficients do not seem to depend much on the state, the only clear difference is between holding cash and holding shares.

The lack of influence of this coefficient is confirmed when one measures the average predicted return conditional on the action of the subjects, which gives results very close to $\omega_0$.

\section{Conclusion}\label{sec:conclusion}

We presented the results of a trading experiment in which the pricing function favours early investment in a risky asset and no posterior trading. In our experimental markets the subjects would make an almost certain gain of over $600 \%$ if they all bought shares in the first period and held them until the end of the experiment. 

However, market impact as defined in Eq.~\ref{eq:impact} acts {\it de facto} as a transaction cost which erodes the earnings of the traders. Our subjects are made well aware of this 
mechanism. Still, when they participate in the experiment for the first time, their trading activity is so high that their profits average to almost zero. They are however found to fare much better when they repeat the experiment as they earn $92 \%$ on average -- which is still much below the performance of the simple risk-neutral rational strategy outlined above. We therefore find that, echoing 
Odean \citep{odean1999tradetoomuch,odeanboys}, {\it investors trade too much}, even in an environment where trading is clearly detrimental and buy-and-hold is an almost certain winning strategy (at variance with real markets where there is nothing like a guaranteed average return of $2 \%$ per period). Our result is potentially quite important when translated into the real world: unwarranted individual decisions can lead to a substantial loss of collective welfare, mediated by the mechanics of financial markets. 

At the end of the experiment we collect data on traders' risk attitude by means of paired lottery choices {\it \`{a} la} \cite{holt2002risk}. We observe that, overall, our subjects are risk-averse and their relative risk aversion increases with the pay-off level in a way that is quantitatively similar to the results reported in \cite{holt2002risk}. We then relate individual risk attitude with the results of the trading experiment and we observe that the activity rate increases and the final wealth decreases with subjects' preference for risk.

Moreover, in each period we also ask subjects to predict the next price of the asset. This provides us with additional information about our controlled experimental market, i.e. we have access not only to the decisions of each trader but also to their expectations. It is important to emphasize that this information would not be available in broker data. In fact, knowing the expectation behind each decision of each trader -- including the decisions to do nothing -- is one of the advantages of laboratory experiments compared to empirical analysis of real data. Using this information, we confirm in Sec.~\ref{ssec:predictions} that the traders in our experiment have a contrarian nature, which, together with the pattern of excessive trading, is one of the known features of individual traders in real financial markets, as discussed in \cite{de2010turnover} and \cite{challet2013robust}. In particular, it seems that our subjects actively engaged in trading in the attempt to ``beat the market'', i.e. trying to make profits by buying the asset when they expected the price to be low and selling the asset when they expected the price to move upwards. 

Contrary to what happens in real financial markets, we have not observed any ``leverage effect'' (i.e. an increase of volatility after down moves). Although there is a clear detrimental collective behaviour in all sessions of this experiment, we do not witness any big crash or avalanche of selling orders that would result
from a panic mode. As we discuss in Sec.~\ref{sssec:collective}, the coordination amongst traders was actually slightly stronger when buying than when selling, resulting in a positive skewness of the returns. In order to induce crashes and study their dynamics, the duration of the experiment could be expanded while the time available for each decision could be reduced. The former would increase the probability of a sudden event by increasing the number of trials and the amounts at stake, while the latter could contribute to higher stress levels amongst subjects and increased sensitivity to price movements. However, we believe that a more efficient way to generate panic and herding behaviour would be to reduce the ``normal'' volatility level while and increasing the amplitude of ``jumps'' in the bare return time series. Within the current setting, 
large fluctuations do not seem surprising enough to trigger panic among our participants. Another idea, perhaps closer to what happens in financial markets, would 
be to increase the impact of sell orders and reduce the impact of buy orders when the price is high, mimicking the fact that buyers are rarer when the price is high (on this point, see the recent results of  \cite{donier2015bitcoin} on the Bitcoin). Other natural extensions of this experiment would include the possibility of fraction orders and hedging, as well as short selling. We leave these extensions and additional experimental designs to future research.

\pagebreak
\newpage
 
\bibliographystyle{apalike}
\bibliography{biblio}

\clearpage
\appendix

\begin{Large}
  \begin{center}
    {\bf Appendix}
  \end{center}
\end{Large}

\section{Additional figures}\label{app:extra_figs}

\begin{figure}[!htpb]
\centering
	\subfloat[Buying orders.\label{fig:c2c3_coll_buy}]{%
		
\includegraphics[scale=0.3]{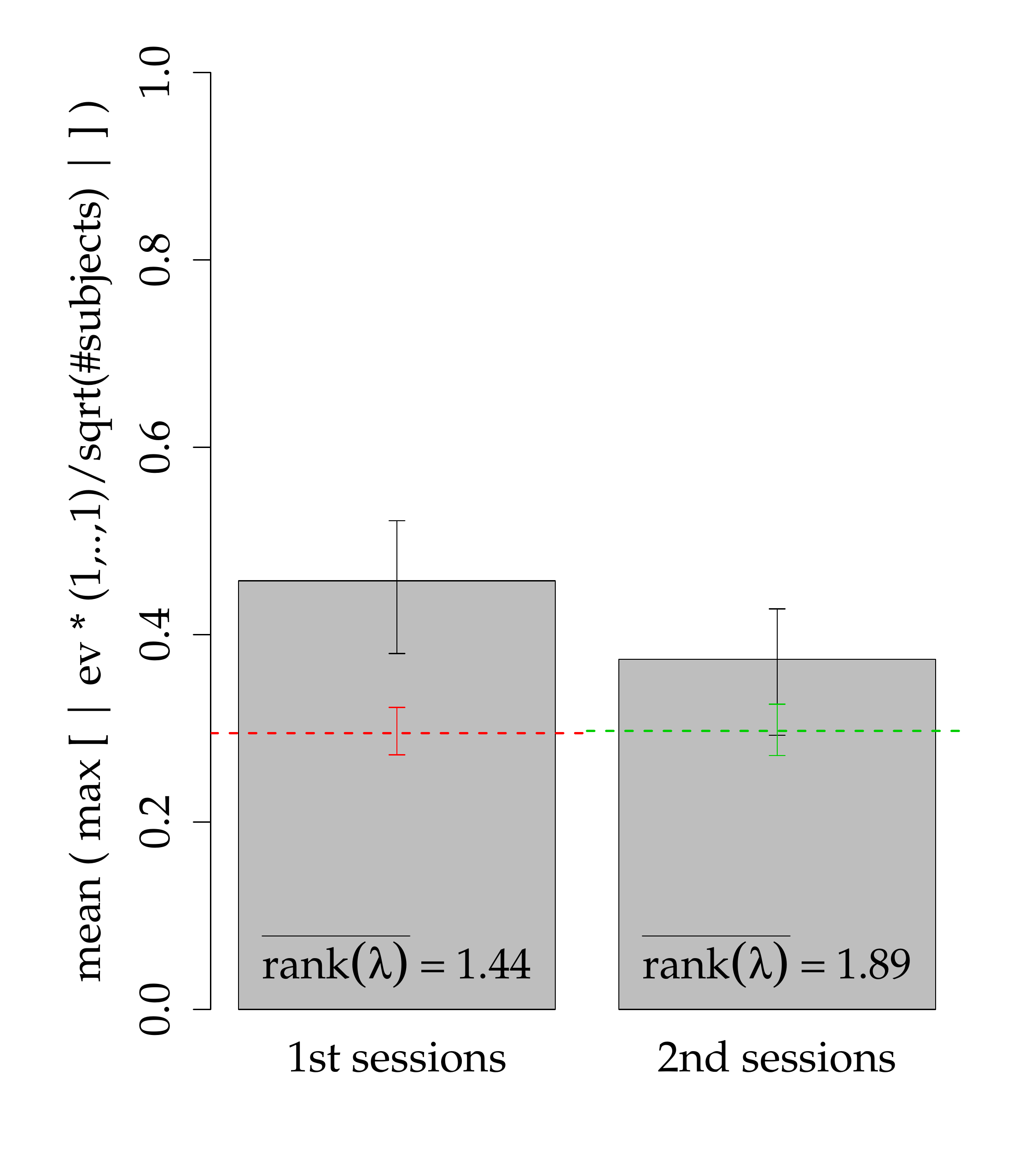}%
		}\hfill
	\subfloat[Selling orders.\label{fig:c2c3_coll_sell}]{%
		
\includegraphics[scale=0.3]{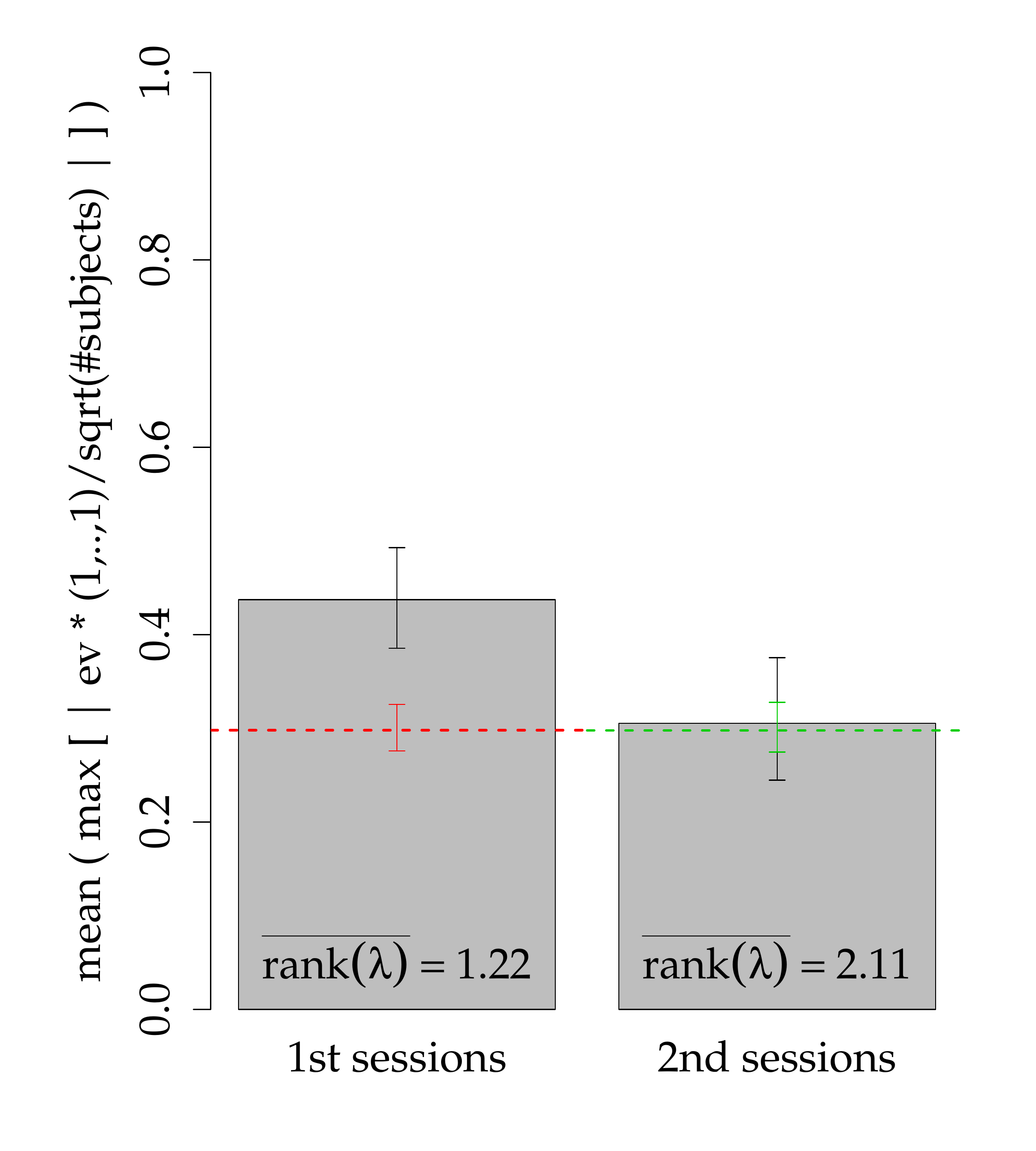}%
	}
	\caption[Content]{Average maximum absolute value of the sum of the 
components of the eigenvectors corresponding to the three largest eigenvalues. 
On the left we restrict ourselves to buying orders, while the result for 
selling orders is shown on the right. The first time step is excluded from the 
data set because we expect a natural bias towards synchronization. The dashed horizontal lines around $0.3$ corresponds to the null hypothesis of completely
uncorrelated actions.}
\end{figure}

\begin{figure}[!htbp] %  figure placement: here, top, bottom, or page
   \centering
   \rotatebox{90}{
   \includegraphics[height=0.45\textheight]{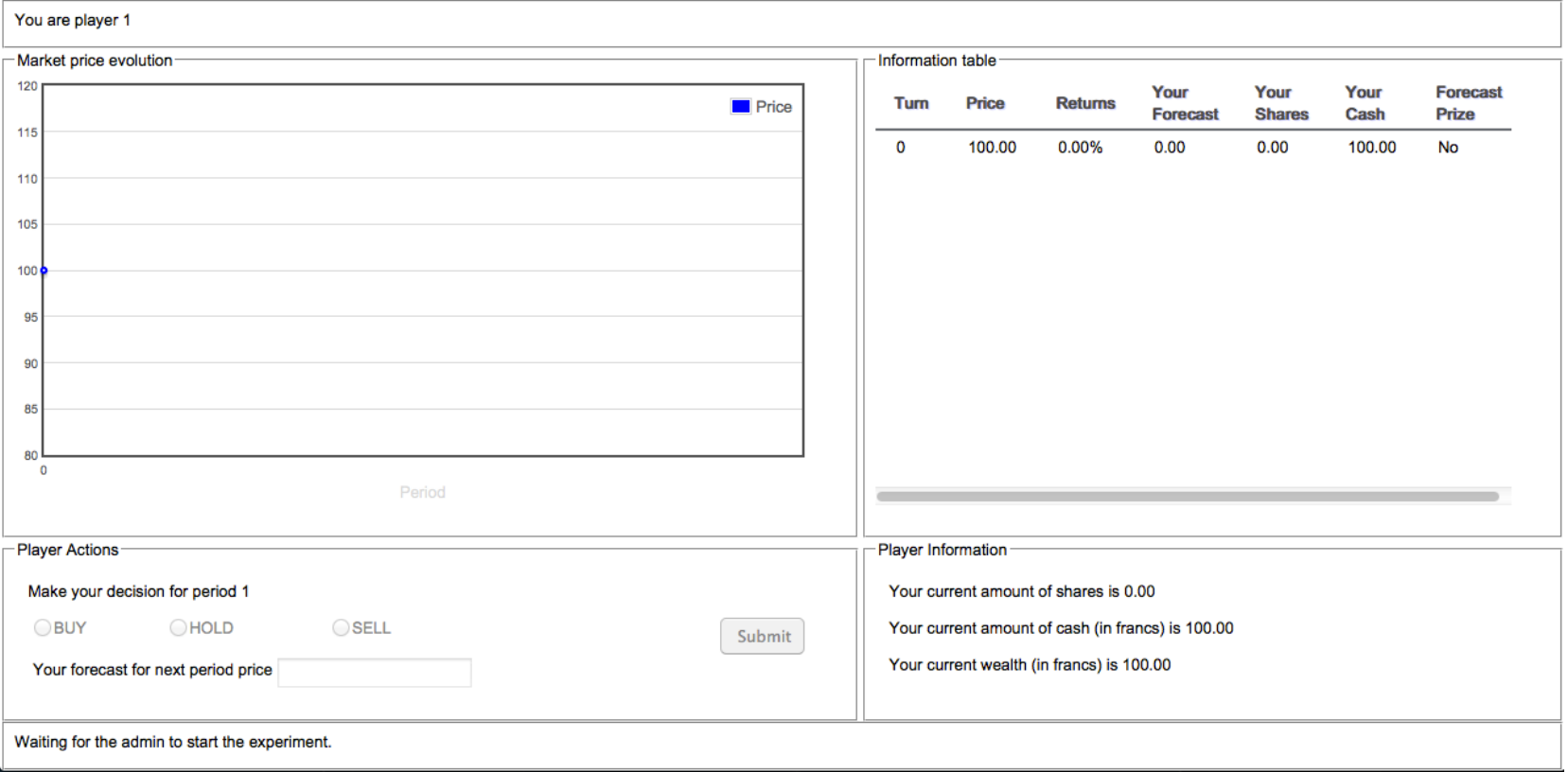}
   }
   \caption{Screenshot}
   \label{fig:screenshot}
\end{figure}

\pagebreak

\section{Estimation of risk aversion parameters}\label{app:risk_est}
As in \cite{holt2002risk}, we use the lottery choices of the subjects to calibrate the utility function described in Eq.~\eqref{eq:utility}. In short, we apply a maximum-likelihood method to find the parameters which maximize the probability that the observed lottery choices are dictated by Eq.~\eqref{eq:utility}.

In this spirit, the first step we take is to model the probability $P_i^R$ that a subject chooses the risky lottery out of the pair $i$ of lotteries. As in \cite{holt2002risk}, we define
\begin{equation}\label{eq:utility_choices}
P_i^R:=\frac{E[U_i^R]^\frac{1}{\mu}}{E[U_i^R]^\frac{1}{\mu}+E[U_i^S]^\frac{1}{\mu}},
\end{equation}
where $E[U_i^R]=\sum_{k=1}^2{p_k U_k^R}$ is the expected utility of the risky lottery and $E[U_i^S]$ is the expected utility of the safe lottery in the pair $i$. Each lottery has two possible outcomes $k=1,2$, each with probability $p_k$ and utility $U_k$ given by Eq.~\eqref{eq:utility}, parameterized by two numbers $\alpha$ and $r$. The parameter $\mu$ is a real number which allows one to consider a range of scenarios between equiprobable choices ($\mu=+\infty$) and utility maximization ($\mu\rightarrow0$). This corresponds to the so called ``logit rule'' (see \cite{bouchaud2013crises} or \cite{anderson1992discrete} for further details).

Secondly, we define the likelihood function
\begin{equation}
\mathcal{L}(\beta, y)=\prod_i (P_i^R)^y_i \cdot (1-P_i^R)^{1-y_i},
\end{equation}
where $y_i$ are the observed choices for each lottery pair, i.e. $y_i=0$ if the subject chose the safe lottery and $y_i=1$ if he chose the risky lottery from pair $i$. In addition, $\beta=[r,\alpha,\mu]$ includes all the model parameters in Eq.~\eqref{eq:utility_choices}.

This way, we have the log-likelihood function
\begin{equation}\label{eq:log-likelihood}
\mathrm{log}[\mathcal{L}(\beta, y)]=\sum_i y_i \mathrm{log}(P_i^R) + (1-y_i)\mathrm{log}(1-P_i^R).
\end{equation}

Finally, the last step is to find the model parameters that maximize Eq.~\eqref{eq:log-likelihood}. We apply the Nelder-Mead algorithm \citep{nelder1965simplex} and use the bias-corrected and accelerated (BCa) bootstrap method \citep{efron1987better} for $95\%$ confidence intervals. The results are summarized in Tab.~\ref{table:util_par}. Quite remarkably, the values of the a-dimensional parameters $r$ and $\mu$ are found to be very close to those reported in \cite{holt2002risk} for their lottery experiments ($\mu = 0.13$, $r=0.27$). In particular, our estimates imply increasing relative risk aversion and decreasing absolute risk aversion.

\begin{table}[!htpb]
\centering
\begin{tabular}{|c|c|c|}
		\hline
		& estimate  & confidence interval \\ \hline
		$\alpha$ & $0.106$  & $[0.085, 0.130]$ \\ \hline
		$r$       & $0.345$ & $[0.263, 0.443]$ \\ \hline
		$\mu$       & $0.114$ & $[0.101, 0.133]$ \\ \hline
\end{tabular}
\caption[Content]{Parameters of Eq.~\eqref{eq:utility_choices} obtained via maximum-likelihood estimation and correspondent $95\%$ confidence intervals. We applied the Nelder-Mead algorithm to maximize Eq.~\eqref{eq:log-likelihood} and used the bias-corrected and accelerated (BCa) bootstrap method for the $95\%$ confidence intervals.}
\label{table:util_par}
\end{table}
%
%\begin{Large}
%  \begin{center}
%    {\bf Appendix (for online publication only)}
%  \end{center}
%\end{Large}

\clearpage
\newpage

\section{Experimental instructions and paired lottery task (for online publication only)}\label{sec:instructions}

%\getpdfpages{./instructions/exp_istr_final_cash.pdf}
%\foreach \x in {1,...,\value{pdfpages}} {
%	
%\frame{\includegraphics[page=\x,scale=0.8]{./instructions/exp_istr_final_cash.pdf}}
%	\clearpage
%}
%\getpdfpages{./instructions/quiz.pdf}
%\foreach \x in {1,...,\value{pdfpages}} {
%	\frame{\includegraphics[page=\x,scale=0.8]{./instructions/quiz.pdf}}
%	\clearpage
%}
%\subsection{Lottery}
%\getpdfpages{./instructions/lottery.pdf}
%\foreach \x in {1,...,\value{pdfpages}} {
%	\frame{\includegraphics[page=\x,scale=0.8]{./instructions/lottery.pdf}}
%	\clearpage
%}


\section*{Supplementary material}

# Experimental Instructions

## 1. Overview

This is an experiment on economic decision making. If you follow the instructions carefully and make good decisions you may earn a considerable amount of money that will be paid to you in cash at the end of the experiment. The whole experiment is computerized, therefore you do not have to submit the paper on your desk. Instead, you can use it to make notes. There is a calculator on your desk. If necessary, you can use it during the experiment. Please do not talk with others for the duration of the experiment. If you have a question please raise your hand and one of the experimenters will answer your question in private.

Today you will participate in one or more market "sequences", each consisting of a number of trading periods. There are two objects of interest in this experiment, *shares* and *cash*, the latter denominated in *francs*. In each period you can trade shares in a market with a computer program, called *market maker*, and the currency used in the market is francs. You pay francs when you buy shares and receive francs when you sell shares. In each period you will have the opportunity to participate in trading or take no action in the market. Details about how this is done are discussed below in section 3. All trading will be in terms of francs. The cash payment to you at the end of the experiment will be in euros. The conversion rate is 4 francs to 1 euro.

## 2. Sequences of trading periods

As mentioned, today's experiment consists of one or more sequences, with each sequence consisting of an uncertain number of periods. Each period lasts 20 seconds. In each period you have to decide if you want to buy shares, sell shares, or hold either your francs or shares, i.e., take no action in the market. The amounts of your shares and francs will be shown on your computer screen. At the beginning of each period a computer program will spin a virtual roulette wheel visualized on your screen colored in blue for a proportion of 99 percent and colored in pink for the remaining 1 percent. If the black pointer ends up in the blue region, the sequence will continue with a new, 20-seconds period. If, instead, the black pointer ends up in the pink region, the sequence will end and your franc balance and the amount of shares for the sequence will be final (see Figure 1). Thus, at the start of each period, there is a 1 percent chance that the period will be the last one played in the sequence, and a 99 percent chance that the sequence will continue with at least one more period.

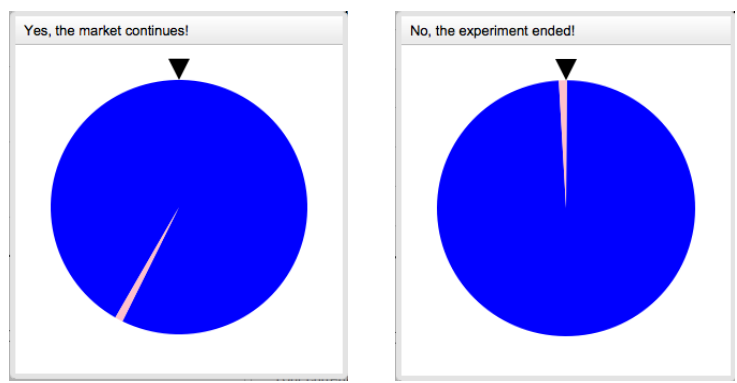

Figure 1. Left panel: experiment continues to next period. Right panel: experiment ends.

If less than 50 minutes have passed since the start of the first sequence, a new sequence will begin. You will start the new sequence just as you started the first sequence. If more than 50 minutes have passed since the beginning of the first sequence then the current sequence will be the last sequence played, meaning that the next time the roulette spin ends up with the black pointer in the pink region the sequence will end and the experiment will be over.

If, by chance, the final sequence has not ended by the three-hour period for which you have been recruited, we will gradually increase the pink region in the roulette wheel until the chance that the sequence will continue equals 0.

**3. Market and trading rules**

The market works as follows. At the beginning of the experiment, each participant will be given an initial endowment of 100 francs. In each period shares can be traded with the market maker. In particular, in each period each participant is allowed to hold either only shares or only francs. Therefore, during each period you may choose to

- Sell <u>all</u> your shares to the market maker in exchange for francs, if you are holding shares;
- Buy shares from the market maker by investing <u>all</u> your francs, if you are holding francs;
- Hold either your shares or francs, and take no action in the market.

Therefore, you cannot buy additional shares if you are already holding shares. Vice versa, you cannot sell shares if you are holding francs.

Trading within each period $t$ occurs according to the following mechanism. First, at the beginning of each period $t$, the computer program spins the roulette wheel. Depending on the outcome of the roulette spin, we can distinguish two cases.

<u>CASE 1.</u> Roulette spin ends up in the blue region: market continues to next period

The price per unit of share in period $t$, denoted by $P_t$, is announced by the market maker and visualized on your computer screen. You can then decide whether to sell shares (if you are holding shares), buy shares (if you are holding francs), or hold either your shares or francs, and take no action in the market. In each period $t$, if you decide to sell or buy, your transaction will take place in the next period at price $P_{t+1}$.

Therefore, if you decide to **sell shares** in period $t$, you will receive an amount of francs in period $t+1$ given by your amount of shares in period $t$ times the price of shares in period $t+1$, which is:

Francs in period $t+1$ = amount of shares in <u>period $t$</u> × price of shares in <u>period $t+1$</u>

If you decide to **buy shares** in period $t$, you will receive an amount of shares in period $t+1$ given by your amount of francs in period $t$ divided by the price of shares in period $t+1$, which is:

Shares in period $t+1$ = amount of francs in <u>period $t$</u> / price of shares in <u>period $t+1$</u>

If you decide to **hold either your shares or your francs** in period $t$, the amount of shares or francs that you own will simply carry over to the next period $t+1$.

Notice that shares need not to be bought or sold in integer units. For example, suppose that in period *t* you own 127.65 francs and you decide to buy shares. Suppose then that the unit price of shares announced in period *t+1* turns out to be 172.50. This means that you will receive 127.65/172.50 = 0.74 shares in period *t+1*.

<u>CASE 2.</u> Roulette spin ends up in the pink region: experiment ends

The market maker announces the price per unit of share in period *t*, denoted by $P_t$, and all the buying and selling orders placed in the previous period *t-1* are executed at price $P_t$ but it will not be possible to make any other decision to buy, sell or hold either shares or francs. Your final market earnings will then be computed as explained in section 4.

The timing of trading is summarized in Figure 2

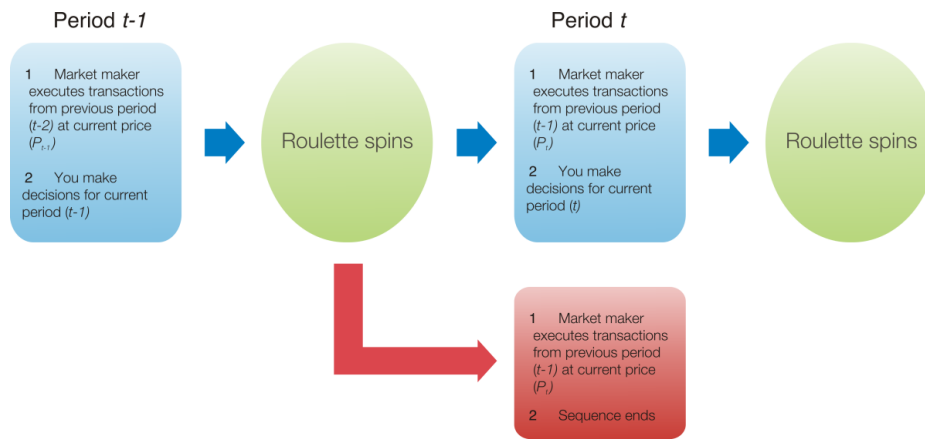

**Figure 2. Timing of trading**

**How the price is determined**

At the end of each period *t*, the market maker will collect the buy and sell orders and use them to determine the price $P_{t+1}$ for the next period *t+1*. The percentage change between the price in period *t+1* and the price in period *t*, also called *return*, is approximately given by the sum of the following components

a) A constant positive term equal to 2%
b) A "trade impact factor" which depends upon the difference between the total amounts of buy and sell orders from the participants in the market. In particular:

- The higher the amount of buy orders in one period, the <u>higher</u> the price in the next period. Therefore, each buy order has a <u>positive</u> impact on price. In particular, if you and all other participants in the market decide to buy shares at the same time, the trade impact factor will be +100%.
- The higher the amount of sell orders in one period, the <u>lower</u> the price in the next period. Therefore, each sell order has a <u>negative</u> impact on price. In particular, if you and all other participants in the market decide to sell shares at the same time, the trade impact factor will be -100%.

Therefore the trade impact factor can take a maximum value of +100% and a minimum value of –100%. If all participants in the market decide to hold their shares or francs, the trade impact factor is zero.

c) Price shocks that can take positive and negative values with the same probability.

Given that the percentage change of prices from one period to the other is approximately given by the sum of the terms listed above (a + b + c), in case of no trading activity by any participant, price grows <u>on average</u> by approximately 2% every period. Figure 3 reports examples of typical price patterns in markets <u>without any trading activity at any period, i.e. if all participants held their shares until the end of the experiment</u>. All numbers in Figure 3 are provided only to give EXAMPLES, they SUGGEST NOTHING about the duration and price realizations of the experiment you are about to start.

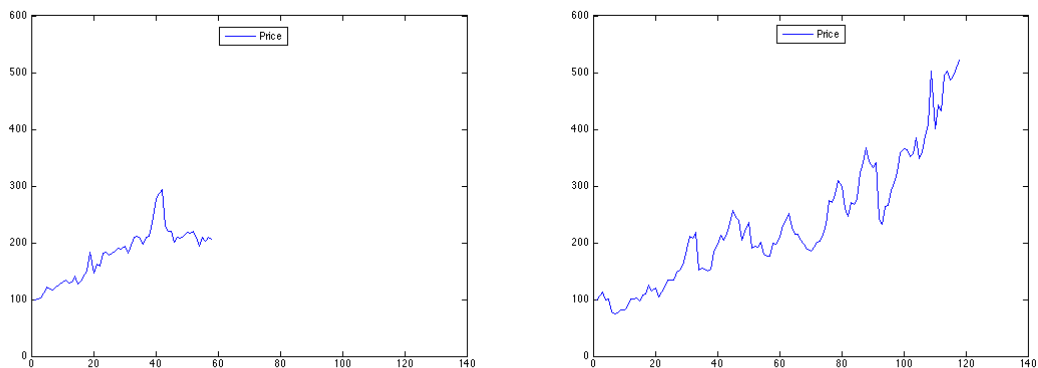

**Figure 3. Examples of price patterns without any trading at any period**

**4. Your earnings**

**Net market earnings**

When a sequence is terminated, that is whenever the roulette spin ends up in pink region, your end-of-sequence balance will be computed.

- If you are holding francs when the sequence is terminated, your amount of francs will determine your end-of-sequence balance.
- If you are holding shares when the sequence is terminated, the market value (in francs) of your shares, given by the amount of your shares times their price in the end of the sequence, will determine your end-of-sequence balance.

Your net market earnings will then be given by your end-of-sequence balance <u>minus</u> 100 francs, corresponding to the initial endowment that you received at the beginning of the experiment:

Net market earnings = end-of-sequence balance – initial endowment.

Therefore your earnings from participating in the market will be given by your end-of-sequence balance in francs minus your initial 100 francs endowment.

**Forecast earnings**

In addition to the money that you can earn from participating in the market, you can earn money by accurately forecasting, in each period $t$, the future price of shares in period $t+1$. You will earn a forecast prize of 0.10 Euro per period if your forecast of the shares' price is within the interval

[0.95 × realized price, 1.05 × realized price].

For example, if the realized price in period $t$ is $p_t$ = 100, you will earn the forecast prize if your forecast for $p_t$ is within the interval [95, 105]. If, for example, the realized price in period $t$ is $p_t$ = 200 you will earn the forecast prize if your forecast for $p_t$ is within the interval [190, 210].

**Total earnings**

Your total earnings for participating in today's experiment will equal the net market earnings that you have at the end of the sequence plus any money that you receive for the forecast task. If your net market earnings are negative, or smaller than the show up fee of 7 Euro, then your earnings from participating in the market will be zero and you will only receive the show up fee of 7 Euro

Earnings from trade = max(net market earnings, show up fee)

plus the money you earned for the forecasting task.

If you participate in more sequences, one of them will be randomly selected and your earnings will equal the total earnings in the selected sequence. As mentioned, the cash payment to you at the end of the experiments will be in Euro. The conversion rate is 4 francs to 1 Euro.

## 5. The computer screen

Below is a sample screen for a fictitious player 1 at the start of period 1.

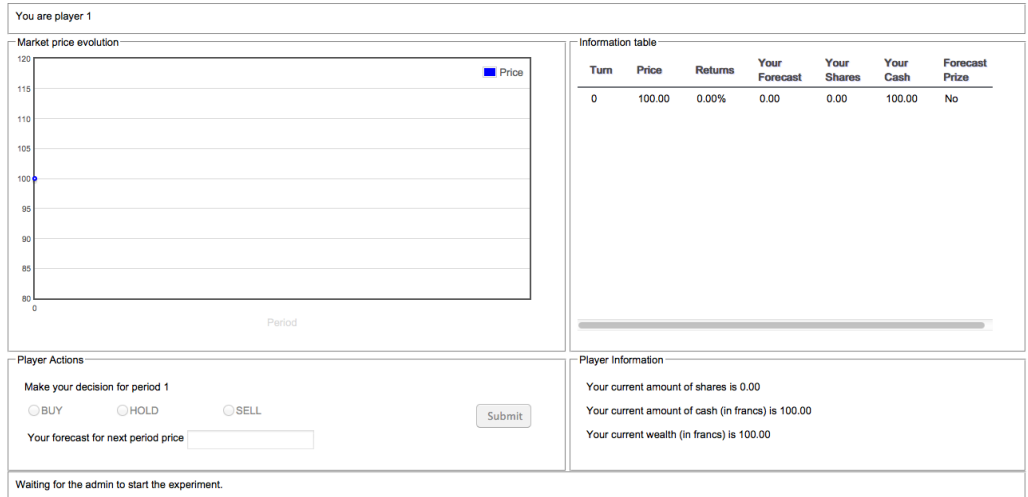

Figure 1. Example of screenshot

In every period, after the roulette spin, each player must perform two tasks:

- Decide whether to buy shares, hold either shares or francs, or sell shares by clicking on the corresponding radio button, i.e., BUY, HOLD, SELL. As explained in section 3, you can decide to buy shares only if in that period you are holding francs, and to sell shares only if in that period you are holding shares. Therefore, in every period, the only active buttons on your screen will be the buttons corresponding to your available actions;
- Enter a forecast of the shares' price in the next period. Please use the dot symbol to separate decimals (example: 10.32).

The box for "Player Actions" is located in the *bottom-left* corner of the screen. After making your choices, you have to submit your decisions by clicking on the "Submit" button.

The box in the *bottom-right* corner of the screen named "Player Information" reports the following information:

- The amount of shares you own in the current period
- The amount of francs you own in the current period
- Your wealth (in francs) in the current period, given by your francs (if you are holding francs) or your shares times the current price (if you are holding shares)

The rest of the screen allows you to track results from previous periods. The graph in the "Market price evolution" box on the *upper-left* corner of the screen reports a graphical representation of the shares' price over time. The table contained in the "Information table" box on the *upper-right* corner of the screen displays additional information about the results in the experiment and it is supplemental to the graph in the left part of the screen. The *first column* of the table shows the time period. The most recent period is always at the top. The *second* and the *third columns* show

respectively the price of shares and the returns, which represent the percentage change in shares price between the current period and the previous. A positive return, say in period 10, means that the price increased from period 9 to period 10, while a negative return in period 10 means that the price decreased from period 9 to period 10. The *fourth column* reports your forecasts (made in the previous period) of the price in the current period. For example, in period 6 the number at the top of the fourth column will report the forecast you entered in period 5 for the price of the shares at period 6. The *fifth* and *sixth columns* report respectively your amounts of shares and francs. Finally, the *seventh column* of the table shows whether in each period you earned the forecast prize or not.

The status bar at the bottom of the screen contains information about the status of the experiment and monitors how much time, out of the 20 seconds constituting the duration of each period, you have to take your decisions. If time is up before you make your choices, the computer program will select HOLD as default action, i.e., you will hold either your shares or francs, and use your previous period forecast.

## 6. Final Quiz

It is important that you understand these instructions. Before continuing with the experiment, we ask that you consider the following scenarios and provide answers to the questions asked in the spaces provided. The numbers used in the quiz are merely illustrative; the actual numbers in the experiment may be quite different. You may find it useful to consult the instructions to answer some of these questions.

**Question 1:** Suppose that a sequence has reached period 25. What is the chance that this sequence will continue with another period, namely period 26? _____. Would your answer be any different if we replaced 25 with 7 and 26 with 8?

- Yes
- No

**Question 2:** Suppose that the sequence has reached period 12, the price announced for that period is $P_{12} = 15$ and you own 5.3 shares. If you decide to sell your shares, how many shares are you allowed to sell?_____. Suppose that you indeed decide to sell in period 12 and that the price announced in period 13 is $P_{13} = 10$. How many francs will you receive?_____.

**Question 3:** Suppose that the sequence has reached period 5, the price announced for that period is $P_5 = 8$ and you own 10 francs. If you decide to buy your shares, how many francs are you allowed to invest?_____. Suppose that you indeed decide to buy in period 5 and that the price announced in period 6 is $P_6 = 20$. How many shares will you receive?_____.

**Question 4:** Suppose that at the beginning of period $t$ a price $P_t = 200$ is announced and then the roulette spin ends up in the pink region. Suppose that you have 1.5 shares. What will be your <u>net</u> market earnings (in francs) at the end of the sequence?_____.

**Question 5:** Suppose that when the experiment ends, i.e., the roulette spin ends up in the pink region, you have 100 francs. What will be your <u>net</u> market earnings (in francs) at the end of the sequence?_____.

**Question 6:** Suppose that in period $t$ you and all other participants decide to hold either your francs or shares and take no action in the market. Do you expect next period's price $P_{t+1}$ to

- Increase
- Decrease

**[Turn sheet for final question]**

**Question 7:** Consider the following scenarios

a) In period *t* you and all other participants have shares. <u>All of you</u> decide to sell.
b) In period *t* you and all other participants have shares. <u>Only you</u> decide to sell.

Regarding scenario a), do you expect next period's price $P_{t+1}$ to

– Increase
– Decrease

How big will the "trade impact factor" (see section "How the price is determined") be in scenario a)? _____ .

Regarding scenario b), do you expect the " trade impact factor" to be:

- Bigger than in scenario a)
- Smaller than in scenario a)

Moreover, do you expect:

- Bigger $P_{t+1}$ in scenario a) than in scenario b)
- Smaller $P_{t+1}$ in scenario a) than in scenario b)

You will now face an additional task that will give you the chance to earn <u>extra money</u>, which will be added to what you already earned in today's experiment.

You will face two sequences (sequence 1 and sequence 2) of 10 decisions each. Details about the decisions that we ask you to make are described in the following sheets.

After you make decisions for both sequence 1 and sequence 2, you will randomly select one of the two sequences by picking a ball from a jar containing balls numbered 1 and 2, and your choice will determine which sequence will be used to determine your payoff. Obviously Sequence 1 and sequence 2 have the same chance of being chosen.

When you have completed all your decisions, and you are satisfied with those decisions please raise your hand and you will be called for payment.

You may now read the instructions on the following sheets.

# SEQUENCE 1

You will face a sequence of 10 decisions. Each decision is a paired choice between two options labeled "Option A" and "Option B". For each decision you must choose either Option A or Option B. After making your choice, please record it on the attached record sheet under the appropriate headings.

The sequence of 10 decisions you will face are as follows:

| Decision | Option A | Option B |
| --- | --- | --- |
| 1 | Receive €4.00 1 in 10 chances OR Receive €3.20 9 in 10 chances | Receive €7.70 1 in 10 chances OR Receive €0.20 9 in 10 chances |
| 2 | Receive €4.00 2 in 10 chances OR Receive €3.20 8 in 10 chances | Receive €7.70 2 in 10 chances OR Receive €0.20 8 in 10 chances |
| 3 | Receive €4.00 3 in 10 chances OR Receive €3.20 7 in 10 chances | Receive €7.70 3 in 10 chances OR Receive €0.20 7 in 10 chances |
| 4 | Receive €4.00 4 in 10 chances OR Receive €3.20 6 in 10 chances | Receive €7.70 4 in 10 chances OR Receive €0.20 6 in 10 chances |
| 5 | Receive €4.00 5 in 10 chances OR Receive €3.20 5 in 10 chances | Receive €7.70 5 in 10 chances OR Receive €0.20 5 in 10 chances |
| 6 | Receive €4.00 6 in 10 chances OR Receive €3.20 4 in 10 chances | Receive €7.70 6 in 10 chances OR Receive €0.20 4 in 10 chances |
| 7 | Receive €4.00 7 in 10 chances OR Receive €3.20 3 in 10 chances | Receive €7.70 7 in 10 chances OR Receive €0.20 3 in 10 chances |
| 8 | Receive €4.00 8 in 10 chances OR Receive €3.20 2 in 10 chances | Receive €7.70 8 in 10 chances OR Receive €0.20 2 in 10 chances |
| 9 | Receive €4.00 9 in 10 chances OR Receive €3.20 1 in 10 chances | Receive €7.70 9 in 10 chances OR Receive €0.20 1 in 10 chances |
| 10 | Receive €4.00 10 in 10 chances OR Receive €3.20 0 in 10 chances | Receive €7.70 10 in 10 chances OR Receive €0.20 0 in 10 chances |

After you have made all 10 decisions, you will be called in a separate room for payment and we will throw a ten-sided die (the faces are numbered from 1 to 10, and the "0" face of the die will serve as 10) twice, once to select one of the ten decisions to be used, and a second time to determine what your payoff is for the option you chose, A or B, for the particular decision selected. Even though you will make ten decisions, only ONE of these will end up affecting your earnings, but you will not know in advance which decision will be used. Obviously, each decision has an equal chance of being used to determine your earnings.

Consider Decision 1. If you choose Option A, then you receive €4.00 if the throw of the ten-sided die is 1, while you receive €3.20 if the throw is 2-10. If you choose Option B, then you receive €7.70 if the throw of the ten-sided die is 1, while you receive €0.20 if the throw is 2-10. The other decisions are similar, except that as you move down the table, the chances of the higher payoff for each option increase. In fact, for Decision 10 in the bottom row, the die will not be needed since each option pays the highest payoff for sure, so your choice here is between €4.00 or €7.70.

Please circle your choice for each of the 10 decisions on your record sheet. Notice that you may choose Option A for some decisions and Option B for others.

# SEQUENCE 2

You will face a sequence of 10 decisions. Each decision is a paired choice between two options labeled "Option A" and "Option B". For each decision you must choose either Option A or Option B. After making your choice, please record it on the attached record sheet under the appropriate headings.

The sequence of 10 decisions you will face are as follows:

| Decision | Option A | Option B |
|---|---|---|
| 1 | Receive €20.00 1 in 10 chances OR Receive €16.00 9 in 10 chances | Receive €38.50 1 in 10 chances OR Receive €1.00  9 in 10 chances |
| 2 | Receive €20.00 2 in 10 chances OR Receive €16.00 8 in 10 chances | Receive €38.50 2 in 10 chances OR Receive €1.00  8 in 10 chances |
| 3 | Receive €20.00 3 in 10 chances OR Receive €16.00 7 in 10 chances | Receive €38.50 3 in 10 chances OR Receive €1.00  7 in 10 chances |
| 4 | Receive €20.00 4 in 10 chances OR Receive €16.00 6 in 10 chances | Receive €38.50 4 in 10 chances OR Receive €1.00  6 in 10 chances |
| 5 | Receive €20.00 5 in 10 chances OR Receive €16.00 5 in 10 chances | Receive €38.50 5 in 10 chances OR Receive €1.00  5 in 10 chances |
| 6 | Receive €20.00 6 in 10 chances OR Receive €16.00 4 in 10 chances | Receive €38.50 6 in 10 chances OR Receive €1.00  4 in 10 chances |
| 7 | Receive €20.00 7 in 10 chances OR Receive €16.00 3 in 10 chances | Receive €38.50 7 in 10 chances OR Receive €1.00  3 in 10 chances |
| 8 | Receive €20.00 8 in 10 chances OR Receive €16.00 2 in 10 chances | Receive €38.50 8 in 10 chances OR Receive €1.00  2 in 10 chances |
| 9 | Receive €20.00 9 in 10 chances OR Receive €16.00 1 in 10 chances | Receive €38.50 9 in 10 chances OR Receive €1.00  1 in 10 chances |
| 10 | Receive €20.00 10 in 10 chances OR Receive €16.00 0 in 10 chances | Receive €38.50 10 in 10 chances OR Receive €1.00  0 in 10 chances |

After you have made all 10 decisions, you will be called in a separate room for payment and we will throw a ten-sided die (the faces are numbered from 1 to 10, and the "0" face of the die will serve as 10) twice, once to select one of the ten decisions to be used, and a second time to determine what your payoff is for the option you chose, A or B, for the particular decision selected. Even though you will make ten decisions, only ONE of these will end up affecting your earnings, but you will not know in advance which decision will be used. Obviously, each decision has an equal chance of being used to determine your earnings.

Consider Decision 1. If you choose Option A, then you receive €20.00 if the throw of the ten-sided die is 1, while you receive €16.00 if the throw is 2-10. If you choose Option B, then you receive €38.50 if the throw of the ten-sided die is 1, while you receive €1.00 if the throw is 2-10. The other decisions are similar, except that as you move down the table, the chances of the higher payoff for each option increase. In fact, for Decision 10 in the bottom row, the die will not be needed since each option pays the highest payoff for sure, so your choice here is between €20.00 or €38.50.

Please circle your choice for each of the 10 decisions on your record sheet. Notice that you may choose Option A for some decisions and Option B for others.

## RECORD SHEET FOR SEQUENCE 1

| | Circle Option Choice |
|---|---|
| Decision 1 | A    B |
| Decision 2 | A    B |
| Decision 3 | A    B |
| Decision 4 | A    B |
| Decision 5 | A    B |
| Decision 6 | A    B |
| Decision 7 | A    B |
| Decision 8 | A    B |
| Decision 9 | A    B |
| Decision 10 | A    B |

# LAB COMPUTER ID:

RECORD SHEET FOR SEQUENCE 2

| | Circle Option Choice |
|---|---|
| Decision 1 | A    B |
| Decision 2 | Circle Option Choice<br>A    B |
| Decision 3 | Circle Option Choice<br>A    B |
| Decision 4 | Circle Option Choice<br>A    B |
| Decision 5 | Circle Option Choice<br>A    B |
| Decision 6 | Circle Option Choice<br>A    B |
| Decision 7 | Circle Option Choice<br>A    B |
| Decision 8 | Circle Option Choice<br>A    B |
| Decision 9 | Circle Option Choice<br>A    B |
| Decision 10 | Circle Option Choice<br>A    B |

# LAB COMPUTER ID:



\end{document}